\documentclass[12pt]{iopart}
\usepackage{graphicx}
\usepackage{color}

\newcommand*\red {\textcolor[rgb]{0.00,0.00,0.00}} %black

\newcommand{\CFM}[0]{{
Centro de F\'{\i}sica de Materiales CSIC-UPV/EHU-Materials Physics Center, E-20018 San Sebasti\'an, Spain}}

\newcommand{\DIPC}[0]{{
Donostia International Physics Center, E-20018 Donostia-San Sebasti\'an, Spain}}
\newcommand{\UPVFA}[0]{{
Departamento de F\'isica Aplicada I,
Universidad del Pa\'is Vasco UPV/EHU,  E-20018 San Sebasti\'an, Spain}}
\newcommand{\IOM}[0]{{
IOM-CNR, Strada Statale 14 Km 163.5, I-34149 Trieste, Italy}}

\usepackage{iopams}
\begin{document}

\title{A ferromagnetic Eu-Pt surface compound grown below hexagonal boron nitride}

\author{Alaa Mohammed Idris Bakhit$^{1,2}$, Khadiza Ali$^{3,4}$, Anna A. Makarova$^5$, Igor P{\'i}{\v s}$^{6}$, Federica Bondino$^6$, Roberto Sant$^7$, Saroj P. Dash$^4$, Rodrigo Castrillo$^{1}$, Yuri Hasegawa$^{1,8}$, J. Enrique Ortega$^{1,2,3}$, Laura Fernandez$^{1}$, and Frederik Schiller$^{1,3}$}

\address{$^1$ \CFM}
\address{$^2$ \UPVFA}
\address{$^3$ \DIPC}
\address{$^4$ Chalmers University of Technology, G{\"o}teborg, Chalmersplatsen 4, 412 96 G{\"o}teborg, Sweden}
\address{$^5$ Physikalische Chemie, Institut f{\"u}r Chemie und Biochemie, Freie Universit{\"a}t Berlin, Arnimallee 22, 14195 Berlin, Germany}
\address{$^6$ \IOM}
\address{$^7$ ESRF, The European Synchrotron, 71 Avenue des Martyrs, CS40220, 38043 Grenoble Cedex 9, France}
\address{$^8$ Department of Physical Sciences, Ritsumeikan University, Kusatsu, 525-8577, Japan}
%\address{$^9$ \UPVFA}
\ead{frederikmichael.schiller@ehu.es}

\begin{abstract}
One of the fundamental applications for monolayer-thick 2D materials is their use as protective layers of metal surfaces and in-situ intercalated reactive materials in ambient conditions. Here we investigate the structural, electronic, and magnetic properties, as well as the chemical stability in air of a very reactive metal, Europium, after intercalation between a hexagonal boron nitride (hBN) layer and a Pt substrate. We demonstrate that Eu intercalation leads to a hBN-{\red{covered}} ferromagnetic EuPt$_2$ surface alloy with divalent Eu$^{2+}$ atoms at the interface. We expose the system to ambient conditions and find a partial conservation of the di-valent signal and hence the Eu-Pt interface. The use of a curved Pt substrate allows us to explore the changes in the Eu valence state and the ambient pressure protection at different substrate planes. The interfacial EuPt$_2$ surface alloy formation remains the same, but the resistance of the protecting hBN layer to ambient conditions is reduced, likely due to a rougher surface and a more discontinuous hBN coating.
\end{abstract}

%
% Uncomment for keywords
\vspace{2pc}

%% main text
\section{Introduction}
\label{Intro}
Ferromagnetic two-dimensional (2D) structures are of uttermost importance in spintronics applications~\cite{Cinchetti2017_NatMat}. Such 2D magnetic systems can either be 2D Van der Waals (vdW) ferromagnets~\cite{gong2017_Nat, huang2017_Nat,Wang2022_ACSNano}, or ultrathin magnetic overlayers~\cite{Gambardella2020_book}. Both types of systems present quantum and topological phases~\cite{Burch2018_Nature}, but achieving new exotic properties requires design and investigation of novel materials and architectures. In thin magnetic overlayers, the surface contains transition and/or rare-earth metals. The interest in such 2D ferromagnetic systems is prompted by the reduced atomic-scale size and the diversity of magnetic states that arise.
Some of the 2D vdW ferromagnets can be exfoliated from bulk crystals, e.g., Fe$_3$GeTe$_2$~\cite{Fei2018_NatureMat} or family members~\cite{Zhang2020_PRB,Chen2022_PRL}. However, most of them are grown by ex-situ chemical vapor or atomic layer deposition~\cite{Geim2013_Nature,Gibertini2019_NatNanotech,Gong2019_Science,Lasek2020_ACSNano}. Magnetic overlayers, on the other hand side, are usually grown by physical vapor deposition under vacuum conditions, a process that does not guarantee an unique 2D layer but may lead to a multilayer system. Achieving a single 2D magnetic layer demands a detailed in-situ, atomic-scale investigation, comprising structure and  electronic states, as well as chemical stability in ambient conditions.
2D magnetic alloys as well as some of the vdW ferromagnets are quite reactive in air, loosing or changing their magnetic properties. Protection of such surfaces can be achieved by means of ceramic coatings, polymer protection films or deposition of non-reactive metals. Nevertheless, most of these protecting overlayers influence the magnetism of the surfaces leading to a variation of the desired properties.
Recently, protection of the surfaces by graphene (Gr)~\cite{Coraux_JPCL2012,Liu2013_NatComm,Martin2015_APL,Weatherup2015_JACS,Cattelan2015_Nanoscale,Naganuma_APL2020,sutter2010_JACS, sokolov2020_MatHor,Anderson2017_PRM}, hexagonal boron nitride (hBN) \cite{Liu2013_NatComm,Caneva2017_ACS_ApplMat,Jiang2017_NanoRes,Tang2021_ACSApplNanoMat,Holler2019_2DMat,Zihlmann2016_2DMat,Ma2022_Nature} and a mixture of both materials~\cite{Pis2018_Carbon} is being considered. In this context, the use of a hBN protecting layer is very appealing, since it would provide close contact of the ferromagnetic material with a wide-gap semiconductor, enabling charge injection. Therefore, the question that arises is whether we can achieve a sufficiently protective hBN layer that preserves the magnetic properties of the 2D compound in ambient conditions.

Here, we study a hBN-protected ferromagnetic Eu-Pt surface alloy. The Eu-Pt compound is formed after Eu intercalation under the hBN film previously grown on a Pt crystal surface.
Metal intercalation below Gr or hBN overlayer has been extensively studied over the last two decades~\cite{sutter2010_JACS, sokolov2020_MatHor, Anderson2017_PRM, Schumacher2014_PRB, Scardamaglia2021_Carbon,Auwarter2019_SSR}. The purpose in the majority of the works was to separate the 2D overlayer from the substrate~\cite{Daukiya2019_ProgSS,Liu2021_JPCC}. Most often this is done by the intercalation of noble metal atoms like Au, Ag, or Cu~\cite{Daukiya2019_ProgSS}. Conversely, in order to force a stronger 2D material interface interaction, one can proceed with the intercalation of alkaline~\cite{Demiroglu2019_JPCL} or earth-alkaline metals~\cite{Grubisic2021_ASS,Kotsakidis2021_AdvMatInter,Kotsakidis2020_ChemMat}. If a too-strongly interacting substrate-2D overlayer is achieved, additional intercalation of oxygen lifts again the 2D layer and re-establishes the original 2D material properties~\cite{sutter2010_JACS}. However, oxygen exposure may result in the oxidation of the protecting hBN layer~\cite{Makarova2019_JPCC}.
Eu intercalation has been less investigated~\cite{Schumacher2014_PRB,schroder2016_2DMat,sokolov2021_JAlloyComp,Anderson2017_PRM,sokolov2020_MatHor}  despite of its interesting magnetic properties, mainly due to the strong reactivity of this rare earth metal.
All Eu intercalation studies have been carried out on graphite or graphene epilayers, but no experiments exist using hBN.

Among the different rare earth metal compounds, europium alloys are particularly interesting due to the various valence states of the Eu atoms. It may adopt a di-valent Eu$^{2+}$, a tri-valent Eu$^{3+}$, or even a mixed-valent state.
For trivalent Eu$^{3+}$, Eu has a 4$f^6$ configuration with $S$ = $L$ = 3 and $J$ = 0. The ground state $J$-multiplet level (in $^{2S+1}L_J$ configuration) is $^7F_0$ presenting a non-magnetic singlet. This situation differs from divalent Eu$^{2+}$ with 4$f^7$ configuration, $S$ = 7/2, $L$ =0 and $J$ = 7/2 leading to a $^8S_{7/2}$ ground state. In the latter case, Eu$^{2+}$ is able to form ferromagnetic compounds, e.g., europium chalcogenides~\cite{McGuire1964_JAP}.
The different valence states of Eu are found to depend on several factors: the surrounding material, the lattice pressure, the number and type of nearest neighbors, etc. In the particular case of Eu-Pt compounds there is a smooth valence transition when changing the stoichiometry from EuPt$_5$ (completely tri-valent) to EuPt$_2$ (Eu atoms in a di-valent state)~\cite{Wickman1968_JPhysChemSol,DeGraaf1980_PhysicaB,Sauer1997_JAllComp}.
Additionally, valence instabilities can be induced in EuPt$_3$ by high pressure~\cite{Ebd-Elmeguid1981_JPC}. Valence changes may also happen at the surface due to a reduced coordination~\cite{johansson79}. Such transitions from trivalent to divalent configuration have also been observed for Eu-Ni or Eu-Pd compounds~\cite{Wieling2002_PRB,Wieling1998_PRB}.

Here we present a Eu-Pt surface alloy formed after intercalation of Eu at the hBN/Pt interface.  First, we perform a structural analysis of the interface, followed by a detailed electronic and magnetic characterization of the Eu-Pt compound. We demonstrate that the topmost layer under the hBN coat is a 2D EuPt$_2$ surface alloy, with di-valent Eu atoms that reveal ferromagnetic behavior at low temperature. Next, we check the efficiency of the hBN layer protection in ambient pressure, by analyzing the electronic properties prior and after air exposure. The sample is a platinum crystal curved around the (111) direction (c-Pt). This provides a smooth variation of the crystallographic orientation across the (macroscopic) surface, allowing us to extend the analysis of the EuPt$_2$ alloy to vicinal Pt crystal planes, characterized by a high density of atomic steps. By scanning our different experimental (electron,photon) probes on top, we can rigorously study the influence of steps and terraces in the structural, magnetic, and electronic properties of the EuPt$_2$ surface alloy, as well as the protecting quality of the hBN layer.

\section{Experimental details}
%--------------------------------------------------------------
\label{Experimental}
The growth and electronic properties were mainly investigated at the Nanophysics laboratory in San Sebastian, Spain using a combined system containing scanning tunneling microscopy (STM), low energy electron diffraction (LEED), X-ray photoemission (XPS) and angle-resolved photoemission spectroscopy (ARPES). Part of the electronic structure investigations have been carried out at BACH beamline of Elettra synchrotron (Trieste, Italy). The XPS setup in the laboratory is equipped with a Specs Al $K_\alpha$  $\mu$-FOCUS 600 monochromator while the ultraviolet light source consists of a Specs UVS-300 discharge lamp with monochromator (Specs TMM 304) tuned to HeII$\alpha$ light with h$\nu$ = 40.8eV. At Elettra synchrotron, $p$-polarized light was applied. All measurements were taken with the sample at room temperature.
STM experiments were carried out in a Omicron VT-setup by holding the sample at room temperature and scanning with a W tip. The analysis of the STM images has been performed with WSXM software~\cite{Horcas2007_RevSciInst}.
The magnetic properties were investigated at ID 32 of the European synchrotron radiation facility (ESRF) by means of X-ray magnetic circular dichroism (XMCD). For this purpose the sample was placed normal or grazing (70$^\circ$) with respect to the incoming photon beam and field. The field was ramped between +6 and -6T with the sample hold at $T$ = 7K. Horizontal, left and right circularly polarized light (99\% polarization) was used for photon energies around the Eu M$_{4,5}$ X-ray absorption edge.

As a substrate material, a cylindrical sector of a Pt (c-Pt) single crystal was used whose cylinder axis is along a [1$\bar 1$0] direction. The centre of the curved surface points towards the [111] direction, while the borders are oriented $\pm$15$^\circ$ with respect to the (111) center (Fig.~\ref{fig:STM}).
This curved Pt surface was cleaned by Ar ion sputtering (room temperature) and temperature annealing (1000K) as well as by occasional oxygen heating (2$\times$10$^{-8}$mbar O$_2$, 950K) followed by a flash in UHV to 1050K. This standard procedure, as described elsewhere~\cite{Walter2015_NatComm,GarciaMartinez2020_AngeChem}, leads to sharp LEED patterns where the typical step splitting was observed.

hBN was grown by chemical vapor deposition (CVD) process from borazine precursor (B$_3$H$_6$N$_3$) (KATCHEM spol. s r.o.). For this purpose the curved Pt crystal was held at 1020K while borazine was dosed for 20 minutes at 2$\times$10$^{-7}$mbar, i.e., for 240L, in order to assure a complete hBN layer. Already a 100L dosage would be enough for a complete layer~\cite{Auwarter2019_SSR}. As can be observed in Fig.~\ref{fig:LEED}, this growth produces a sharp and well ordered moir{\'e} pattern in LEED at the Pt(111) position of the curved crystal. On the other substrate positions, corresponding to the vicinal surfaces in the mentioned $\pm$15$^\circ$ range around (111), the LEED reveals less ordered structures with line like features pointing to a multi-facet structure.

Eu was deposited in a third step on top of this hBN/c-Pt substrate while the sample was held at an elevated temperature to allow Eu intercalation below hBN. As pointed out earlier~\cite{Schumacher2014_PRB}, the high temperature is quite important to immediately protect the Eu from oxidation. We used substrate temperatures between 570 and 870K. The deposition process was carried out in UHV systems with a base pressure prior to Eu deposition below 1$\times$10$^{-9}$mbar not surpassing 3$\times$10$^{-9}$mbar during deposition. For lower substrate temperatures incomplete intercalation takes place and part of the Eu stays on top of the hBN, see supplementary material for details.
The oxidation protection experiments consisted in exposing the sample to ambient pressure conditions (6 hours, room temperature, 80\% humidity).
The Eu thickness given throughout the manuscript corresponds to the values obtained for the calibration of the evaporator with a quartz microbalance at the measurement position. Since the sticking coefficient for both the quartz microbalance and the sample are unknown, but probably different, the indicated numbers can be compared with each other but are not suitable for absolute values.

\section{Results and Discussion}
\label{Results}
\subsection{Formation of Eu-Pt surface alloy below hBN}
\subsubsection{Eu intercalation in the hBN/Pt(111) interface}
\begin{figure*}[b!]
%\centerline{\includegraphics[width=1.0\columnwidth]{LEED_evolution_new.pdf}}
\centerline{\includegraphics[width=1.0\columnwidth]{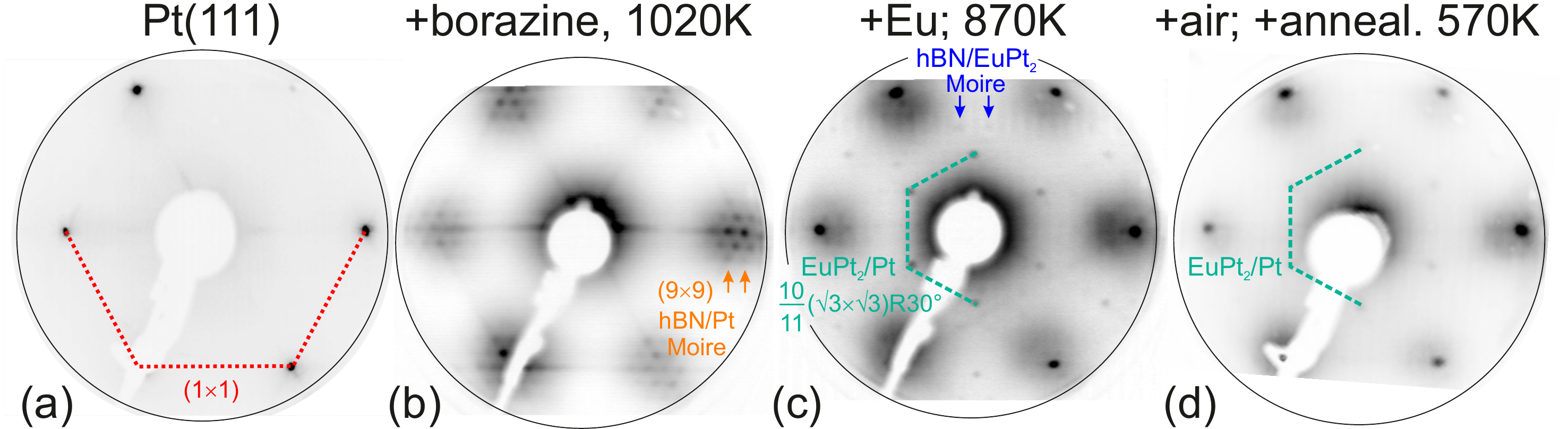}}
\caption{\textbf{LEED images along the preparation process.}
(a): Pt(111), (b) after borazine exposure at 1020K producing a (9$\times$9) moir{\'e} pattern, (c) additional Eu deposition/intercalation at $T_{sample}$ = 870K leading to EuPt$_2$ layer below hBN and a different moir{\'e} pattern, (d) after 6 hours room temperature air exposure and subsequent 570K vacuum annealing (LEED kinetic energy 70eV).}
\label{fig:LEED}
\end{figure*}

The structural evolution of the hBN/Eu/Pt intercalated system can readily be monitored with low energy electron diffraction (LEED) experiments. LEED patterns in Figure~\ref{fig:LEED} correspond to the (111) plane on the Pt curved crystal. The clean Pt(111) pattern is shown in Fig.~\ref{fig:LEED}(a), which transforms, after borazine dosing at $T$ = 1020K, into the characteristic hBN (9$\times$9) moir{\'e} of Fig.~\ref{fig:LEED}(b)~\cite{PREOBRAJENSKI2007_PRB}. For successful Eu intercalation, there is a threshold temperature of the substrate of $T$ = 770K. Below this temperature part of the rare-earth material stays on top of the hBN coat, being subject to rapid contamination/oxidation (see Supplementary Information). At the lower range of the complete intercalation temperatures, right above 770K, the LEED pattern only shows the progressive extinction of the hBN moir{\'e}. When rising the temperature to $T$ = 870K a new $\approx$($\sqrt{3}\times\sqrt{3})$R30$^\circ$ pattern emerges [Fig.~\ref{fig:LEED}(c)], with some weak satellite spots. A detailed inspection of the $\approx$($\sqrt{3}\times\sqrt{3})$ structure reveals a 10/11$\cdot$($\sqrt{3}\times\sqrt{3})$R30$^\circ$ geometry with respect to Pt(111). This pattern reflects the presence of a EuPt$_2$/Pt(111) moir{\'e}-like coincidence lattice, similar to those found in rare earth RE-Au and RE-Ag surface alloys with RE-Au$_2$ and RE-Ag$_2$ composition~\cite{Corso2010_PRL,Corso2010_ACSNano,Ormaza_PRB2013,Fernandez2020_Nanoscale,Xu2020_PCCP,Que2020_JPCL,Ormaza_NanoLett2016}. The ($\sqrt{3}\times\sqrt{3}$) ordering arises from the 1:2 Pt:Eu stoichiometry of the alloy at the local atomic-scale.
The pre-factor emerges from the lattice mismatch of the 2D RE-noble metal alloy layer and the substrate.

The hBN/Eu/Pt(111) system is different from the Gr/Eu/Ir(111) one~\cite{Schumacher2014_PRB}, where the superstructure LEED spots belong to the graphene diffracted beams. In that case it was proposed that Eu forms a floating layer between the Ir(111) substrate and the graphene layer. In the here considered hBN/Eu/Pt system, however, the strongest LEED spots correspond to the EuPt$_2$ layer at the Pt interface. As indicated in Fig.~\ref{fig:LEED}(c), we can still detect extra satellite spots around the 10/11$\cdot$($\sqrt{3}\times\sqrt{3})$R30$^\circ$ structure, which arise from the coincidence lattice defined by the mismatched hBN/EuPt$_2$ interface. Being all LEED structures properly identified, one can calculate real space lattice parameters out of the pattern. Taking into account the Pt lattice constant of $a_{Pt}$ = 3.92\AA, we obtain the EuPt$_2$ coincidence lattice constant of 11$\cdot$$a_{Pt}$/$\sqrt{2}$ = 30.5\AA, from which we deduce the EuPt$_2$ lattice parameter $a_{EuPt_2}$ = $a_{Pt}$/$\sqrt{2}\cdot$10/11$\cdot\sqrt{3}$ = 5.29\AA, with a nearest neighbor distance of 3.05\AA. The lattice mismatch of the  EuPt$_2$ layer with the hBN lattice on top (2.504\AA) is quite large, but explains the 4.6$\times$4.6 weak superstructure spots, marked by blue arrows in Fig.~\ref{fig:LEED}(c). After 6 hours of air exposure and a soft annealing to 570K to remove air adsorbates the LEED pattern still reveals the 10/11$\cdot$($\sqrt{3}\times\sqrt{3})$R30$^\circ$ structure.

\begin{figure}[tb!]
%\centerline{\includegraphics[width=1.0\columnwidth]{Structural_Model_new.pdf}}
\centerline{\includegraphics[width=1.0\columnwidth]{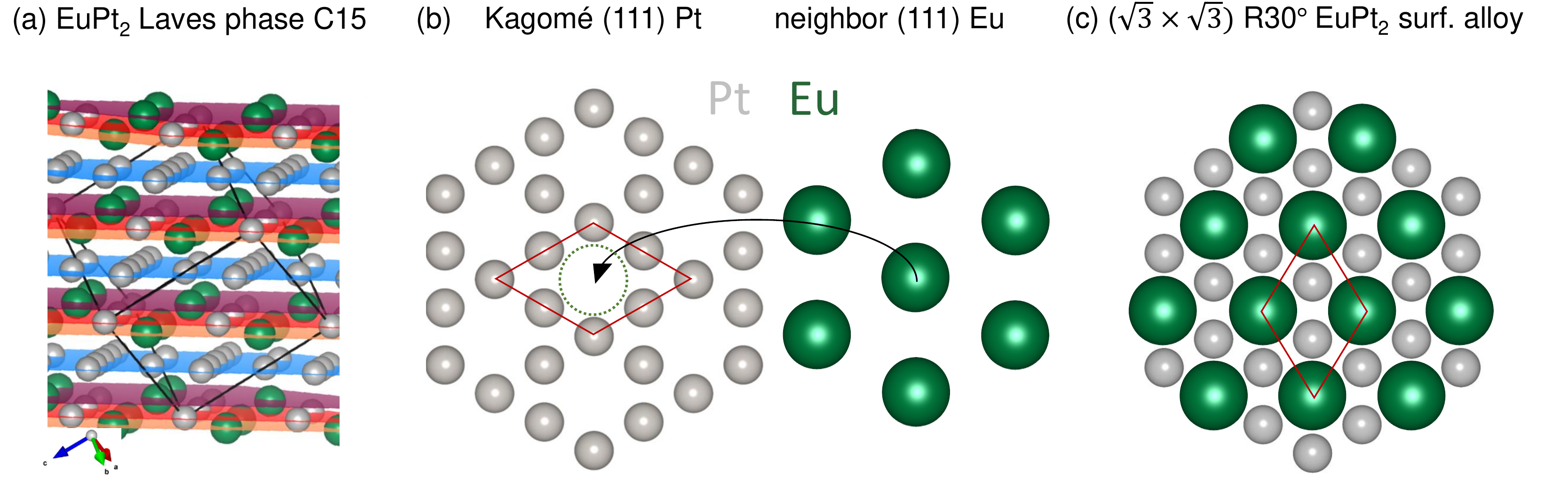}}
\caption{\textbf{Structural Model of the EuPt$_2$ surface.}
(a): Laves phase (C15) of bulk EuPt$_2$ in the MgCu$_2$ structure, (b) central Pt Kagom{\'e} (blue layer in (a)) and neighboring Eu (111) layer stacking inside bulk EuPt$_2$ that would result in a PtEu$_3$ composition with (2$\times$2) superstructure after Eu incorporation, (c) ($\sqrt{3}\times\sqrt{3}$)R30$^\circ$ monolayer surface alloy structure below a 2D hBN layer.}
\label{fig:Structure}
\end{figure}

Bulk EuPt$_2$ exists and crystalizes in a MgCu$_2$ Laves phase structure, as shown in Fig.~\ref{fig:Structure}(a). A bulk lattice constant $a_{Laves}$ between 7.64 and 7.73\AA\ has been reported~\cite{Wickman1968_JPhysChemSol,Erdmann1973SolStChem,DeGraaf1980_PhysicaB}. However, in such bulk EuPt$_2$ structure, and along the [111] direction, one cannot find any stoichiometric EuPt$_2$ plane. Contiguous (111) planes contain  either Pt and Eu solely, as shown in Fig.~\ref{fig:Structure}(b). The plane containing the Eu atoms is shifted by 3/8 of the densely packed Pt planes (approx. 0.8\AA).
The fundamental Pt containing (111) plane is formed by a Kagom{\'e} lattice [blue layer in Fig.~\ref{fig:Structure}(a)], with the Eu atoms sitting above and below each Pt hexagon of the Kagom{\'e} lattice. Considering the Eu-Pt bilayer, this defines a (2$\times$2) superstructure with a EuPt$_3$ composition. The here found EuPt$_2$ structure formed below hBN is therefore not related to bulk EuPt$_2$ and can be understood by the simple incorporation of Eu atoms in the uppermost Pt(111) surface plane.
Due to the larger size of the Eu atom, the interatomic distance at the surface increases, and a mismatch with the Pt(111) substrate underneath arises, leading to the 10/11$\cdot$($\sqrt{3}\times\sqrt{3})$R30$^\circ$ coincidence lattice.

\begin{figure}[tb!]
%\centerline{\includegraphics[width=0.5\columnwidth]{Eu3d_sev_prep.pdf}}
\centerline{\includegraphics[width=0.5\columnwidth]{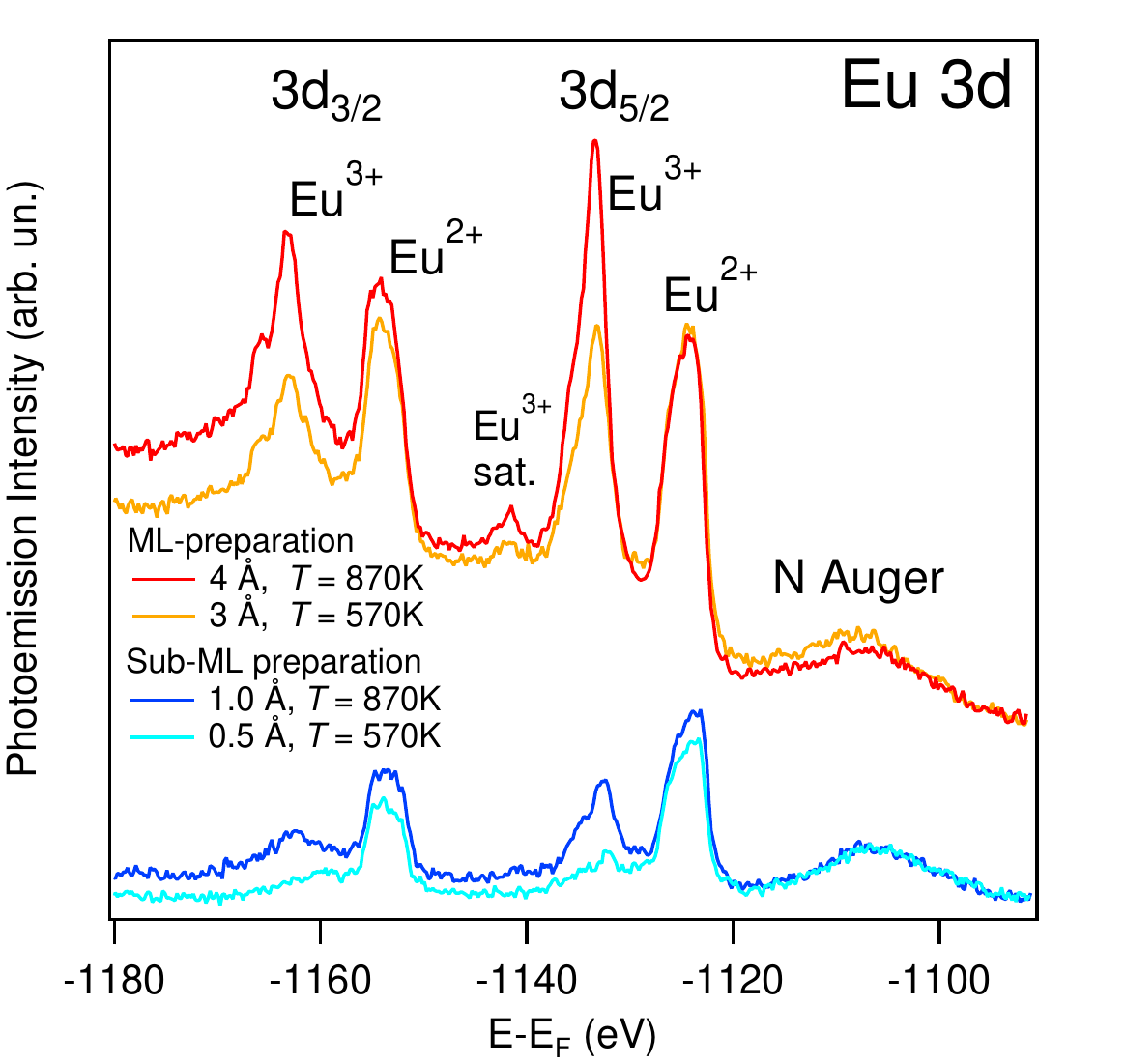}}
\caption{\textbf{Eu 3d photoemission signal.}
X-ray photoemission spectra for the Eu 3d edge taken at h$\nu$ = 1486.6eV (Al K$_\alpha$) for sub- and monolayer preparations at sample temperatures $T$ = 570K and 870K, respectively.}
\label{fig:XPS_Eu3d}
\end{figure}

The chemical characterization of the Eu intercalation process is carried out by means of X-ray photoemission spectroscopy. Results are shown in Fig.~\ref{fig:XPS_Eu3d}. In the bulk EuPt$_2$ compound Eu atoms are in a di-valent Eu$^{2+}$ configuration, while Eu atoms in compounds with higher Pt content become mixed-valent or completely tri-valent~\cite{DeGraaf1980_PhysicaB}. The latter situation would be the case for Eu interstitial atoms, e.g., those that diffuse into the bulk and are surrounded by Pt completely. Fig.~\ref{fig:XPS_Eu3d} reveals the Eu 3d core level for sub- and monolayer preparations at different temperatures. Submonolayer Eu deposition and intercalation at low substrate temperature ($T$ = 570K) leads to a (nearly) complete divalent configuration while preparations at higher $T$ result in the appearance of an additional Eu$^{3+}$ signal. We interpret these observations as follows: at lower temperature only a partial Eu intercalation takes place, leading to the formation of EuPt$_2$ patches below hBN (intercalated) and di-valent metallic Eu above hBN (not intercalated). For higher temperature, the Eu intercalation is complete, but together with the Eu$^{2+}$ species of the EuPt$_2$ interface below hBN, Eu$^{3+}$ component arises, which is ascribed to Eu interstitials in the Pt bulk, or simply to the buildup, under the topmost EuPt$_2$ patches, of EuPt$_n$ ($n>$ 2) alloys, giving rise to a tri-valent or a mixed-valent situation. {\red{The Eu$_2$O$_3$ oxide has also Eu atoms in a tri-valent state. However, the presence of significant amounts of Eu$_2$O$_3$  are ruled out by the absence of an O 1s emission in XPS. The spectral region of the latter can be found in the supporting material.}} At higher coverage ($>$3 \AA), preparations at both low and high temperatures already force extra Eu atoms to diffuse below the completed EuPt$_2$ layer, leading to similar di- and tri-valent contributions in the Eu 3d XPS spectra, as shown in the top part of Fig.~\ref{fig:XPS_Eu3d}. This tri-valent contribution for low-temperature preparation prove that a considerable part of Eu intercalates already at this temperature. Increasing the temperature at high coverage enhances the bulk diffusion, and leads to even stronger Eu$^{3+}$ emission compared to Eu$^{2+}$.

\subsubsection{Eu intercalation in vicinal hBN/Pt(111) interfaces}

\begin{figure*}[tb!]
\centerline{\includegraphics[width=1\columnwidth]{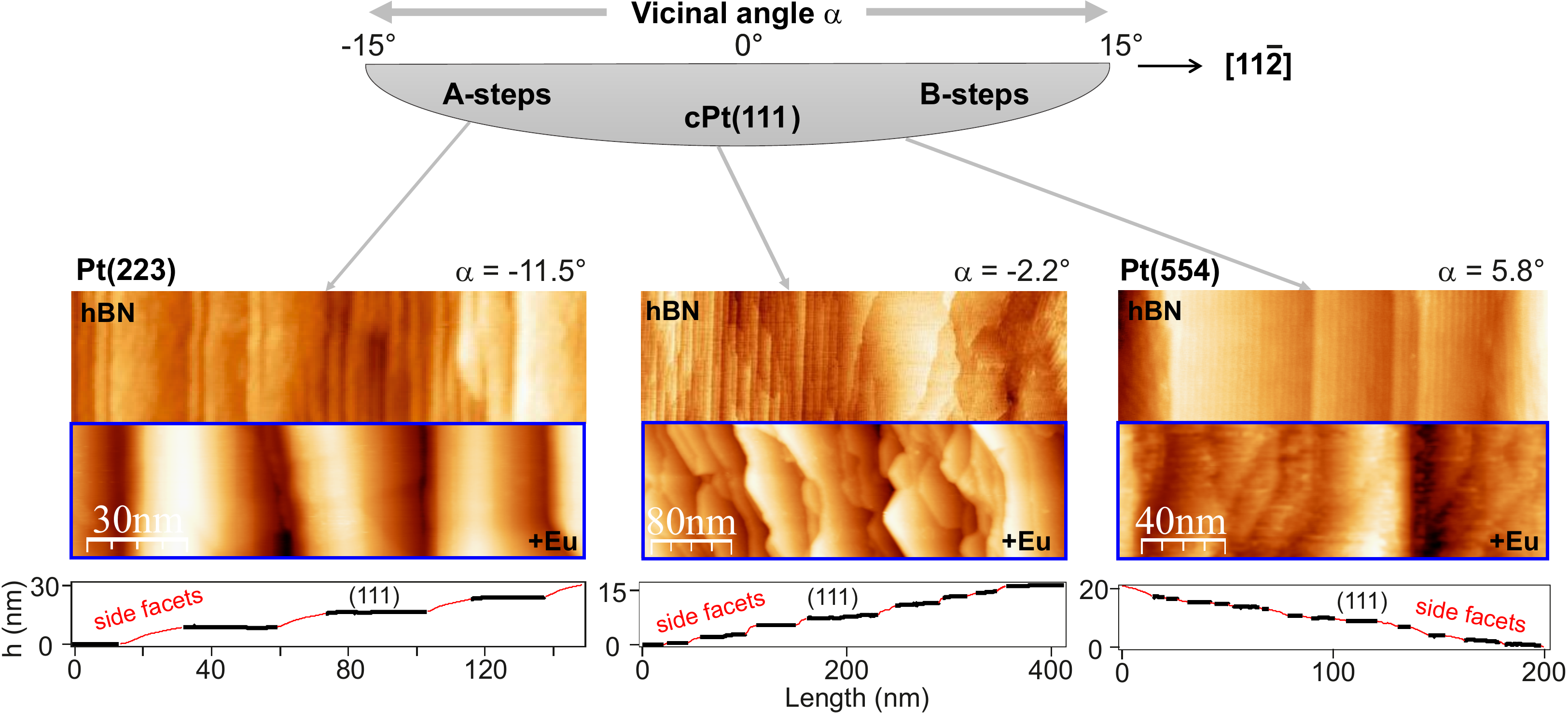}}
%\centerline{\includegraphics[width=1\columnwidth]{STM_fred_with_3LineScans.pdf}}
\caption{\textbf{STM images of the (Eu)/hBN/c-Pt(111) system.}
Large scale STM images of a hBN monolayer prior and after Eu intercalation collected at three different positions on the curved c-Pt(111) substrate (scanning parameters: I = 0.13nA; U = 0.5V). The line scans were taken close to the top of the images corresponding to the Eu intercalated systems. (111) and side facets are indicated by different color.}
\label{fig:STM}
\end{figure*}
After characterizing the Eu intercalation below the hBN monolayer on Pt(111), we focus on surfaces vicinal to the Pt(111) plane, investigated with the curved sample sketched in Fig.~\ref{fig:STM}. The negative sign of the vicinal angle $\alpha$ corresponds to surfaces with A-type steps (\{100\} microfacets), and the positive to B-type step arrays (\{111\} microfacets)~\cite{Walter2015_NatComm}. STM images in Fig.~\ref{fig:STM} correspond to three representative points of the curved substrate, namely the Pt(223) position ($\alpha$ = -11.5$^\circ$), a low vicinal angle ($\alpha$ = -2.2$^\circ$) close to Pt(111), and the Pt(554) surface ($\alpha$ = 5.8$^\circ$). Prior to hBN growth, all vicinal surfaces exhibit well-ordered 1D step arrays, either at low and high vicinal angles~\cite{Walter2015_NatComm}.
However, the hBN monolayer induces drastic structural changes, leading to a more complex nanoscale landscape. Close to the (111) position, large hBN/Pt(111) areas develop, which alternate with densely bunched steps. At larger vicinal angles the step bunching process remains, and the surface becomes a faceted structure. At the (554) position one observes a rather well ordered structure consisting of (111) terraces and side facets tilted at approx. 12$^\circ$. At the (223) position one does not get a clear long range order. The latter situation is similar to stepped Ni crystals covered by hBN~\cite{Fernandez2019_2DMat}, while the ordered structures are rather close to the Rh case, where hBN growth on stepped surfaces leads to periodically arranged nanofacets~\cite{Ali2021_SciAdv}.
One interesting question is whether the hBN forms a continuous, defect-free coat over the hill-and-valley structure underneath, since this requires ``bending'' of the hBN layer
over Pt substrate facets.
Due to step-bunching/faceting, the hBN monolayer must bend at the facet borders, i.e., step edges of the underlying Pt substrate must be overgrown by hBN to create a defect-free film. Such a defect-free connection of adjacent terraces has been shown for hBN growth on Cu(110)~\cite{Wang2019_Nature} and Cu(111)~\cite{Chen2020_Nature}. In the former case, the closure of the film was attributed to the fact that the Cu(110) step height $s$ of $s$ = 1.27\AA\, is smaller than the hBN bonding length (1.44 \AA)~\cite{Bets2019_NL}. However, for Cu(111) with $s$ = 2.08\AA\, this explanation alone would not allow for ``carpeting''. Nevertheless, the edge-docking probability of the B$_6$N$_7$ seed to the different A- or B-type steps was observed to be different~\cite{Chen2020_Nature} resulting in a crystalline film. For Pt(111) with an even higher step height $s$ = 2.27\AA\, this connection ability is unknown. For a similar material, Ru(0001), a continuous film is difficult to achieve due to the intrinsic mismatch between individually nucleated h-BN domains on the same terrace as well as between adjacent terraces~\cite{Lu2013_JACS}.
In general terms, however, one expects an increasing number of defects for an increasing presence of facet/step boundaries at large vicinal angles.

Eu deposition and intercalation on the vicinal surfaces changes the facet periodicity, size and inclination, as observed in the STM images. At the A-step type Pt(223) position, the rather disordered hBN/Pt(223) structure transforms into a well ordered array after Eu intercalation. On B-type steps, however, the Eu intercalation is leading to smaller facets. A statistical analysis of the STM images reveals an increasing average facet distance [(111)+side facet] of 20 facets/$\mu m$ close to (111), 25 facets/$\mu m$ at (223), and 65 facets/$\mu m$ at (554). This means that at A-type steps the  surface rugosity decreases, contrary to B-type steps. On the other hand, XPS spectra (see below) reveal a similar Eu$^{2+}$/Eu$^{3+}$ relation on the different crystal positions, with a slightly higher di-valent amount at (111) compared to stepped surface  planes.

\subsection{Magnetism of the hBN/EuPt$_2$ system in the (111) plane}
\begin{figure}[tb!]
%\centerline{\includegraphics[width=0.5\columnwidth]{MagProp_Eu_hBN_Eu_Pt111_new.pdf}}
\centerline{\includegraphics[width=0.5\columnwidth]{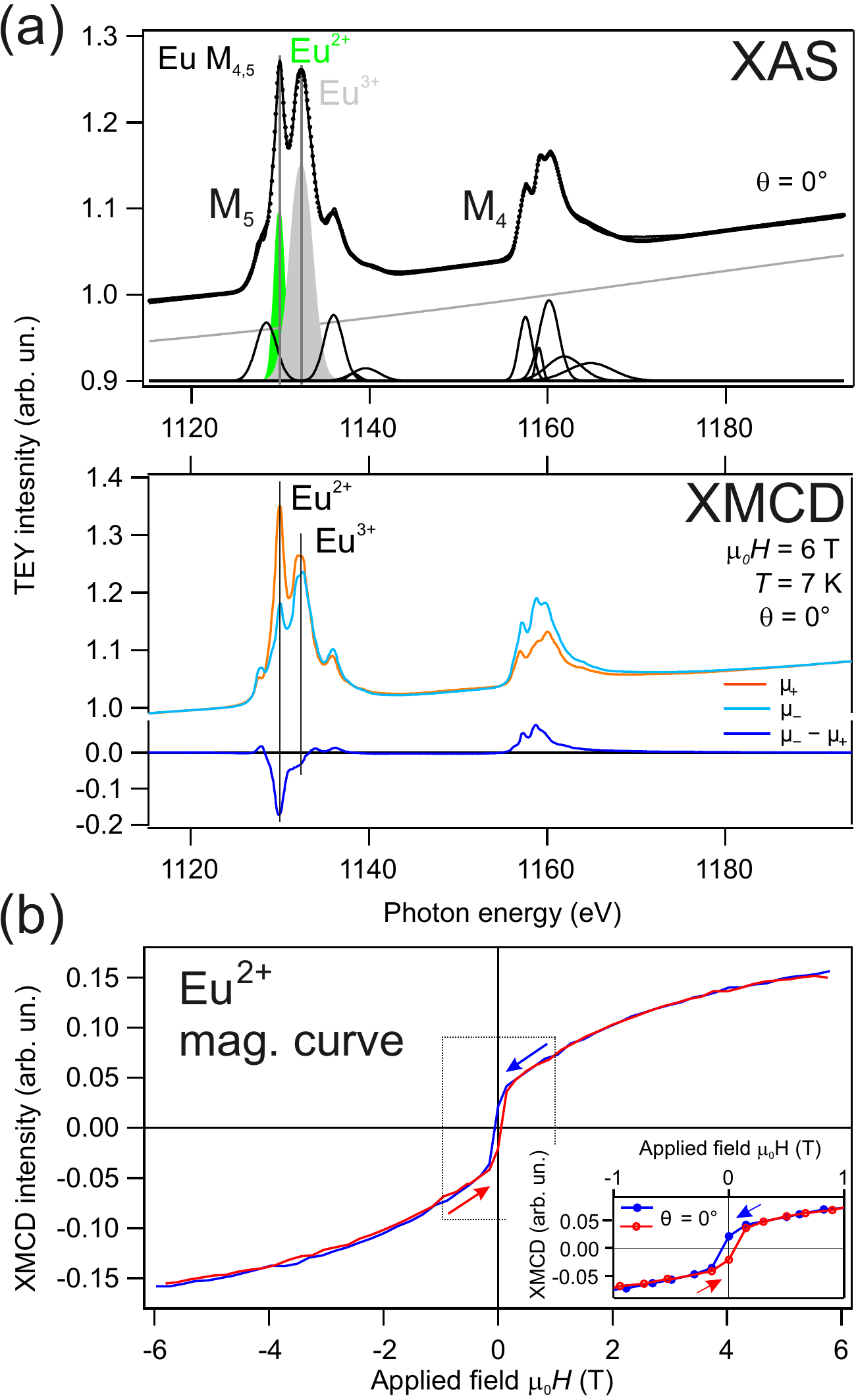}}
\caption{\textbf{Magnetic properties of intercalated Eu below hBN/Pt(111).}
(a) X-ray absorption spectrum (total electron yield - TEY) of horizontally (top, XAS) and circularly polarized light with opposite sign and its resulting difference spectrum shown below (XMCD) of 3\AA~Eu intercalated below hBN/Pt(111) at a substrate temperature of 770K. The XAS spectrum was fitted with several gaussian profiles (black, grey, green) and a linear background (grey line) for each of the two main Eu$^{2+}$ and Eu$^{3+}$ contributions. (b) Magnetization curve taken at the maximum of the Eu M$_{5}$ XMCD signal for a variable applied field from  +6T to -6T (blue) and in the opposite direction (red), respectively. The sample was oriented perpendicular to the applied field at temperature $T$ = 7K. The inset presents a zoom-in to the small field region.}
\label{magnetism}
\end{figure}
Divalent Eu (4$f^7$,  $J$ = 7/2) has a low-temperature ferromagnetic state in bulk EuPt$_2$ and similar compounds~\cite{Nakamura2016_JPhysSocJap}. We measured the magnetic properties of our hBN-protected EuPt$_2$ surface alloy with X-ray circular magnetic dichroism (XMCD) at the (111) position in the curved crystal. Results are shown in Fig.~\ref{magnetism}. The X-ray absorption (XAS) spectrum in part (a) reveals a mixture of Eu$^{2+}$ and Eu$^{3+}$ contribution after Eu intercalation below the hBN/Pt(111) surface. This observation confirms the coexistence of the two Eu valences and it is consistent with the XPS results, namely di-valent Eu atoms at the EuPt$_2$ surface and tri-valent Eu atoms diffused into the Pt bulk. The XMCD signal results from the difference of the absorption spectra of left and right circularly polarized light and is shown in the bottom of Fig.~\ref{magnetism}(a). At the applied field of 6 T, it shows the typical lineshape of pure di-valent Eu~\cite{Blanco2022_PRRes}. This is expected since, as stated before, Eu$^{3+}$ has a 4$f^6$ configuration with $S$ = $L$ = 3 and $J$ = 0, hence the Eu$^{3+}$ signal should not contribute significantly to the anisotropy.

Ferromagnetism can be probed by measuring the Eu XMCD signal while varying the applied magnetic field. The XMCD signal is proportional to the magnetization $M$ in the system. The resulting magnetization curve is shown in Fig.~\ref{magnetism}(b). It reveals a ``S'' shape-like behaviour with
some jump close to zero-field.
Ferromagnetism is confirmed by the small hysteresis opening, which is better observed in the inset of Fig.~\ref{magnetism}(b). The corresponding XMCD spectrum after removing the applied field (remanent state) is shown in the supplementary material. Another method for determining the ferromagnetic state is the Arrott plot analysis~\cite{Arrott1957_PR}, which is also presented in the supplementary material. For the in-plane geometry, with the sample at 70$^\circ$ with respect to the magnetic field and light incidence, the magnetization curve is even more ``S''-shaped at small fields, pointing to an out-of-plane easy axis (see supplementary material for more details).

%Such a value is expected for temperatures close to the Curie temperature.

\subsection{Exposure of the hBN/Eu/Pt system to air}
\begin{figure}[tb!]
%\centerline{\includegraphics[width=0.5\columnwidth]{XPS_onPt111.pdf}}
\centerline{\includegraphics[width=0.5\columnwidth]{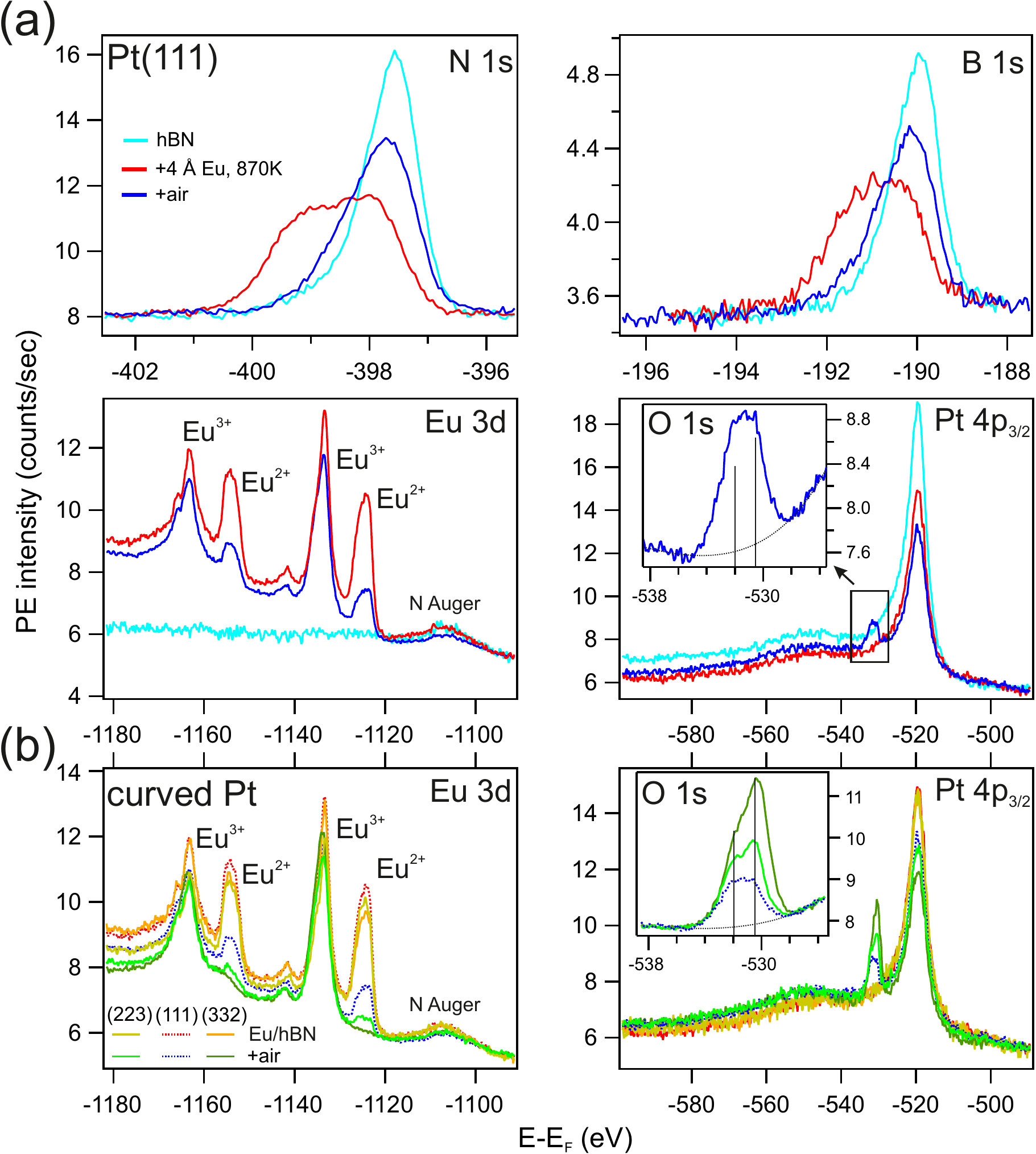}}
\caption{\textbf{XPS analysis of Eu intercalation below hBN on the curved Pt crystal.}
(a) N 1s, B 1s, Eu 3d, O 1s, and Pt 4p$_{3/2}$ X-ray photoemission spectra taken at h$\nu$ = 1486.6eV (Al K$_\alpha$) for the Eu 4\AA~preparation before and after exposing to ambient conditions at Pt(111). (b) Comparative Eu 3d, O 1s, and Pt 4p$_{3/2}$ core levels at three positions of the curved Pt substrate, at Pt(332), Pt(111) and Pt(335) positions, respectively.}
\label{fig:XPS_allPt111}
\end{figure}
Fig.~\ref{fig:XPS_allPt111}(a) displays the complete evolution of the XPS spectra during the intercalation of 4-\AA-Eu in the hBN/Pt(111) interface and after exposure to air. Prior to Eu intercalation, we obtain the characteristic shape and positions in the B 1s and N 1s core levels for the weakly-coupled hBN/Pt(111) system~\cite{PREOBRAJENSKI2007_PRB}. {\red{Individual contributions of the emissions can be extracted from line fittings, that are shown with further details in the supplementary material.}} After Eu intercalation B 1s and N 1s core levels notably change their shape and energy, reflecting the fact that the hBN contact interface is now different. {\red{There are emissions from areas where the hBN layer is still contacting to Pt without Eu intercalation, giving rise to the right hand side emission, and areas for hBN on EuPt$_2$, to which we assign the additional peaks that appear shifted to high binding energy. Such a shift is also observed for the hBN/Rh(111) interface, pointing to a similarly strongly  interacting hBN/PtEu$_2$ interface.~\cite{PREOBRAJENSKI2007_PRB}}} With respect to the Eu core level, as mentioned above, it proves that Eu atoms are present in two configurations, di-valent Eu$^{2+}$ for the 2D EuPt$_2$ alloy and tri-valent Eu$^{3+}$ for Eu atoms incorporated into the Pt bulk below the EuPt$_2$ layer. After sample exposition to ambient conditions (6 hours, room temperature, 80\% humidity) and a vacuum annealing to 770K to remove at least part of impurities from the air exposure, the ratio of Eu$^{2+}$/Eu$^{3+}$ drops to one third. Note, that also the other core level intensities shrink. This relates to the appearance of adventitious C 1s (see supplementary material) and O 1s emissions. The O 1s core level is detected at the higher binding energy side of the Pt 4p$_{3/2}$ spectrum. A detailed view (inset) indicates a double-peak with energy positions of 530.5eV and 532eV, respectively. The former is in good agreement with the binding energy in the tri-valent Eu$_2$O$_3$ oxide~\cite{Mercier2006_JElecSpecRelPhen,Baltrus2019_SurfSciSpec}. The 532eV emisssion is interpreted as due to hydroxide -OH, typical of Eu samples exposed to air. On the other hand, B 1s and N 1s core levels partially recover their shape and energy prior to Eu exposure. {\red{Especially the very strong interacting component of the hBN/EuPt$_2$ areas disappeared while the emissions of weak interacting hBN/Eu and hBN/Pt are preserved or even increased. The latter suggests that hBN areas on oxidized Eu have a similar core level emission as the hBN/Pt system, being both very low interacting interfaces. Another result from the XPS analysis is that}} oxidation of the hBN layer is not observed.

The hBN protection for the intercalated EuPt$_2$ alloy in vicinal Pt substrates is examined in Fig.~\ref{fig:XPS_allPt111}(b), in direct comparison with the Pt(111) plane. Here we show the Eu 3d, Pt 4p$_{3/2}$ and O 1s core levels at the Pt(332) and Pt(223) surfaces. At the Pt(332) plane, no di-valent signal remains in the Eu 3d spectrum after exposure to air. At the Pt(223) position the Eu$^{2+}$/Eu$^{3+}$ is strongly reduced, but the Eu$^{2+}$ peak is still visible. Again, the O 1s spectrum suggests that the strong reduction of the number of di-valent Eu atoms is due to the formation of tri-valent Eu oxides and hydroxides. The overall O 1s intensity increases as the Eu$^{2+}$ peak decreases at the stepped surfaces, although a more detailed peak analysis indicates that the intensity of the Eu-OH peak at 532eV is similar in all three cases, and it is the 530.5eV peak from Eu$_2$O$_3$ the one that scales reciprocally with the Eu$^{2+}$/Eu$^{3+}$ ratio.
The complete oxidation of the B-type (332) surface correlates with the high facet density observed in the STM analysis of Fig.~\ref{fig:STM}, since this allows a higher number of hBN bending or breaks at facet borders. The rather small divalent signal that remains in Pt(223) may reflect the presence of larger (111) and side facets where the intercalated EuPt$_2$ alloy remains better protected at ambient conditions.

\begin{figure}[tb!]
%\centerline{\includegraphics[width=0.8\columnwidth]{AirExpos_effect_EuhBNPt111_allhor.pdf}}
\centerline{\includegraphics[width=0.8\columnwidth]{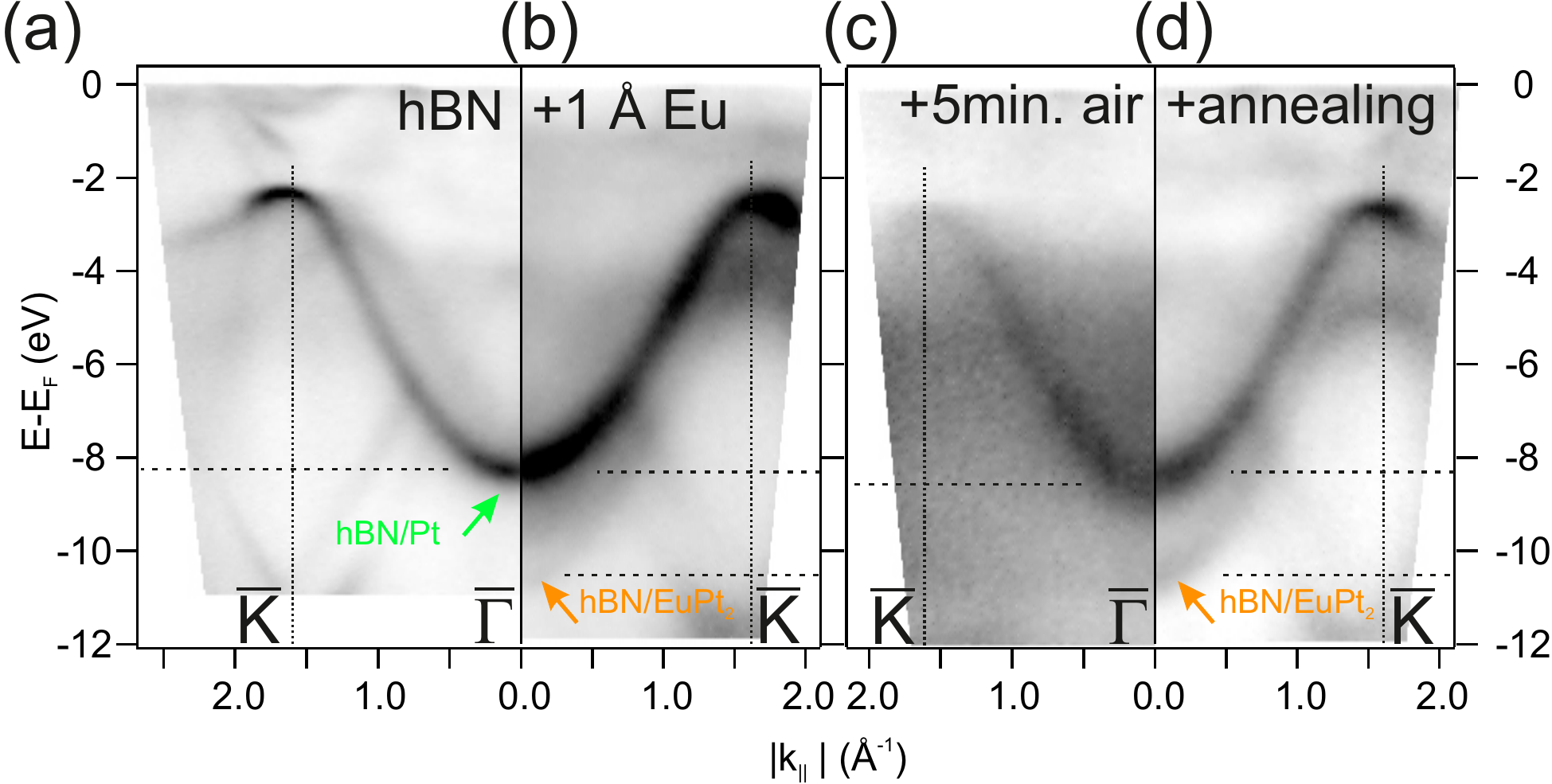}}
\caption{\textbf{Angle resolved Photoemission analysis of the hBN layer.}
He IIa (h$\nu$ = 40.8eV) photoemission intensity maps along $\bar\Gamma\bar K$ direction of the hBN band structure for (a) hBN/Pt(111), (b) after intercalation of 1 \AA~ of Eu at $T$ = 770K, (c) exposure of that sample to air, and (d) after another annealing to 770K to desorb air contamination, respectively. The most intense features correspond to the $\pi$-band of hBN at the indicated interfaces.}
\label{fig:air_exposure_bands}
\end{figure}
Finally, we analyze the impact of the air exposure on the hBN protecting layer. For this purpose angle-resolved photoemission spectroscopy (ARPES)  is particularly appropriate, since it is highly sensitive to the hBN valence band, as shown in Fig.~\ref{fig:air_exposure_bands}. The photoemission intensity map of Fig.~\ref{fig:air_exposure_bands}(a) corresponds to the hBN/Pt(111) interface (centre of the curved sample), and has been measured using the He II excitation energy (h$\nu$ = 40.8eV). The strongly dispersing, intense features are the hBN $\pi$ bands, with minimum at the $\bar\Gamma$ point of the Brillouin zone and maximum at $\bar K$. The emission between approx. 2 and 8 eV binding energy entirely belongs to hBN, while the Pt valence band features appear closer to the Fermi level. After intercalation of a very small amount of Eu (1 \AA ), the main $\pi$ band appears unaltered, although a replica of this band emerges below, at 2 eV higher binding energy (Fig.~\ref{fig:air_exposure_bands}(b)). The unaltered band corresponds to {\red{hBN $\pi$-band emissions of}} pure Pt areas without Eu while the shifted band arises in areas where the Eu intercalates. The band shift is explained by the enhanced interaction of the hBN with the EuPt$_2$ substrate, with a net electron transfer from Eu atoms to the hBN layer similar to other systems~\cite{schroder2016_2DMat,Larciprete2012_ACSNano,Jolie2014_PRB,Pervan2015_RPB}. After air exposure all bands get quite blurry, due to adsorbates.
The adsorbate layer is disordered, which has the consequence that substrate and interface emissions are not well diffracted, decreasing the band intensity and increasing the background in second electron emissions. The dominating hBN $\pi$ band is still visible, slightly shifted to higher binding energies. Interestingly, both the pure hBN/Pt and the Eu-intercalated bands can be recovered by annealing the sample again to 770K, which removes adsorbates from the surface. {\red{First, this indicates that the oxygen exposure does not affect (oxidize) the hBN layer, which remains intact. Second, the photoemission intensity mapping still includes the higher binding energy replica, indicating that at least part of the strongly interacting hBN/EuPt$_2$ patches remain, despite the oxidation of another part of the EuPt$_2$ layer as observed in the XPS. Third, above the oxidized part, the pi-band of the hBN layer is expected to present an energy position similar to the original hBN/Pt(111) $\pi$ band position since the charge transfer from the metallic Eu has disappeared and the interaction at hBN/Pt or hBN/oxide interfaces is weak in both cases. This leads to two very similar $\pi$-bands that cannot be resolved. In the supporting material we present data from the $\bar\Gamma$-point emission of an preparation of 4\AA\ Eu where this effect is better observed.}}

\section{Conclusions}
We have investigated the structural, magnetic and electronic properties of Eu after intercalation between a Pt substrate and a hBN monolayer. We used a Pt sample curved around the (111) direction in order to additionally assess the role of substrate steps. Our LEED analysis of the (111) interface shows a $\sim$($\sqrt{3}\times\sqrt{3}$)R30$^\circ$ pattern, revealing the presence of the EuPt$_{2}$ surface alloy under the hBN layer. We find that Eu atoms in this EuPt$_{2}$ layer are divalent, while Eu atoms that have diffused further inside the Pt bulk during the intercalation process are trivalent. Interestingly, the Eu$^{2+}$/Eu$^{3+}$ ratio is not affected by the presence of steps at the Pt substrate. XMCD magnetization curves on the di-valent Eu atom reveal a ferromagnetic behavior. Air exposure of the sample leads to a partial protection of the divalent Eu atoms at the (111) plane, while at vicinal surfaces the protecting role of the hBN layer is less efficient, as reflected in the larger attenuation of the divalent Eu state. Such incomplete protection of vicinal planes may be related to a larger number of defects and domain boundaries in a more discontinuous hBN layer, since this covers a much rougher hill-and-valley faceted structure. This facilitates oxygen diffusion, intercalation and the EuPt$_{2}$ alloy oxidation. In contrast, the hBN layer itself remains intact upon both Eu intercalation and air exposure.

\section*{Conflicts of interest}
There are no conflicts to declare.

\section*{Statement of Contributions}
A.M.I.B, K.A., and F.S. conceived and designed the work. A.M.I.B., K.A., A.A.M., I.P., F.B., L.F., and F.S. carried out structure and photoemission data collection while the data were analyzed and interpreted by A.M.I.B, K.A., and F.S. The magnetic measurements were performed by A.M.I.B., R.S., R.C., Y.H., L.F. and F.S. The article was drafted by A.M.I.B., J.E.O., and F.S. All authors contributed in the critical revision of the article and its final approval.

\section*{Supplementary material}
%The Supporting Information is available free of charge.
Supplementary material contains information on electron spectroscopy analysis for Eu intercalation at insufficient substrate temperatures as well as additional XPS spectra for the Eu intercalation and air exposure process. Furthermore, for the case of magnetic properties investigations using XMCD technique, the remanent XMCD spectrum and the Arrot plot analysis is presented. Additionally, the magnetic easy axis direction is investigated.

%\section*{Acknowledgement}
\ack
We acknowledge financial support from grants PID2020-116093RB-C44 funded by the Spanish MCIN/AEI/ 10.13039/501100011033 and the Basque Government (Grant  IT-1591-22). We acknowledge the European Synchrotron Radiation Facility for provision of beam time on ID32. ESRF access was provided through proposal MA-5454~\cite{ESRF_Ma5454}. Part of the research leading to the result has been supported by the project CALIPSOplus under Grant Agreement 730872 from the EU Framework Programme for Research and Innovation HORIZON 2020.
Y. H. appreciates the support of Japan Society for the Promotion of Science (JSPS) Overseas Research Fellowships and
I.P. and F.B. acknowledge financial support from EUROFEL project (RoadMap Esfri).

\pagebreak

%\section{TOC figure}
%\begin{figure*}[htb]
%\centerline{\includegraphics[width=\columnwidth]{Fig1inrto.pdf}}
%\centerline{\includegraphics[width=\columnwidth]{TOC.pdf}}
%\caption{\textbf{Exposure of a ferromagnetic surface alloy of EuPt$_2$ below a protection layer of hBN towards ambient conditions.}
%}
%\label{fig:intro}
%\end{figure*}

%\bibliographystyle{iop_articleTitles}
%\bibliography{../bib}

\end{document}

\title[Supplementary material]{Supplementary material of ``A ferromagnetic Eu-Pt surface compound grown below hexagonal boron nitride''}

\author{Alaa Mohammed Idris Bakhit$^{1,2}$, Khadiza Ali$^3,4$, Anna A. Makarova$^5$, Igor P{\'i}{\v s}$^{6}$, Federica Bondino$^6$, Roberto Sant$^7$, Saroj P. Dash$^4$, Rodrigo Castrillo$^{1}$, Yuri Hasegawa$^{1,8}$, J. Enrique Ortega$^{1,2,3}$, Laura Fernandez$^{1}$, and Frederik Schiller$^{1,3}$}

\address{$^1$ \CFM}
\address{$^2$ \UPVFA}
\address{$^3$ \DIPC}
\address{$^4$ Chalmers University of Technology, G{\"o}teborg, Chalmersplatsen 4, 412 96 G{\"o}teborg, Sweden}
\address{$^5$ Physikalische Chemie, Institut f{\"u}r Chemie und Biochemie, Freie Universit{\"a}t Berlin, Arnimallee 22, 14195 Berlin, Germany}
\address{$^6$ \IOM}
\address{$^7$ ESRF, The European Synchrotron, 71 Avenue des Martyrs, CS40220, 38043 Grenoble Cedex 9, France}
\address{$^8$ Department of Physical Sciences, Ritsumeikan University, Kusatsu, 525-8577, Japan}
\ead{frederikmichael.schiller@ehu.es}

%\begin{suppinfo}

\renewcommand{\thepage}{S\arabic{page}}
\renewcommand{\thesection}{S\arabic{section}}
\renewcommand{\thetable}{S\arabic{table}}
\renewcommand{\thefigure}{S\arabic{figure}}
\setcounter{figure}{0}
\setcounter{page}{1}

The supplementary information file contains:

Supplementary Figures~\ref{XAS}-\ref{mag_loops}

Supplementary information and data are provided to further illustrate the intercalation process of Eu below the hBN/Pt substrate as well as magnetic easy axis determination of the Eu-Pt alloy below hBN.

\pagebreak

\section{Spectroscopy analysis of insufficient temperature for Eu intercalation}
Eu does not completely intercalate below a hBN layer on Pt if the substrate temperature is not sufficiently high. Here we will show Near-edge X-ray absorption fine structure (NEXAFS) and X-ray photoelectron spectroscopy (XPS) data for such preparations. Data were taken at ID32 and BACH beamlines of ESRF and Elettra synchrotrons, respectively.

\begin{figure}[b!]
\centerline{\includegraphics[width=0.75\textwidth,angle=0,clip]{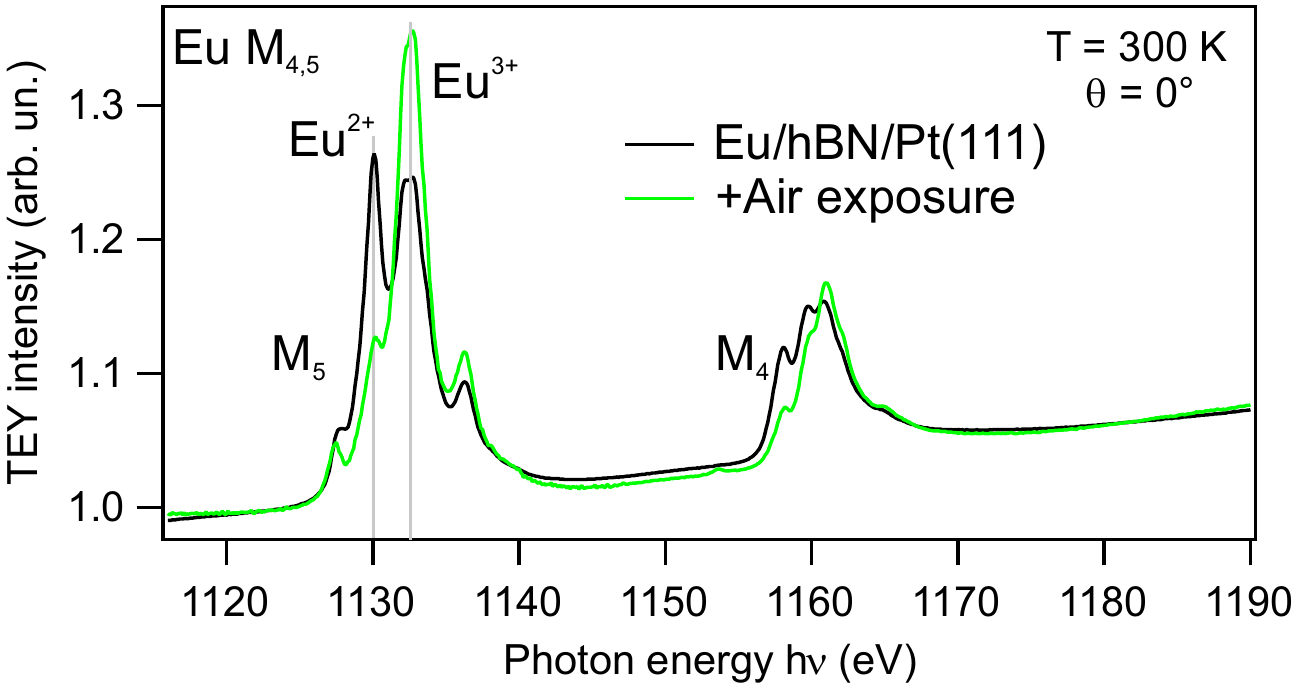}}
\caption{\textbf{X-ray absorption spectra at the Eu M$_{4,5}$ absorption edge for Eu deposition on hBN/Pt(111) at $T$ = 770K and successive air exposure. Measurement carried out at room temperature and with the sample at normal incidence.}}
\label{XAS}
\end{figure}
X-ray absorption spectroscopy (normal to the field and light incidence) for a approx. 1 ML thick Eu film deposited at a $T$ = 770K hot hBN/Pt(111) substrate (10 minutes) and successive exposure to air is shown in Fig.~\ref{XAS}. The air exposure was carried out with the sample at room temperature during five minutes. After air exposure a strong transformation of the spectral shape towards tri-valent Eu configuration took place. Nevertheless, even after the air exposure experiment a small di-valent part is still present. In order to remove remaining air adsorbates, the sample was post-annealed in UHV to $T$ = 470K during 10 minutes.

\begin{figure}[t!]
\centerline{\includegraphics[width=0.75\textwidth,angle=0,clip]{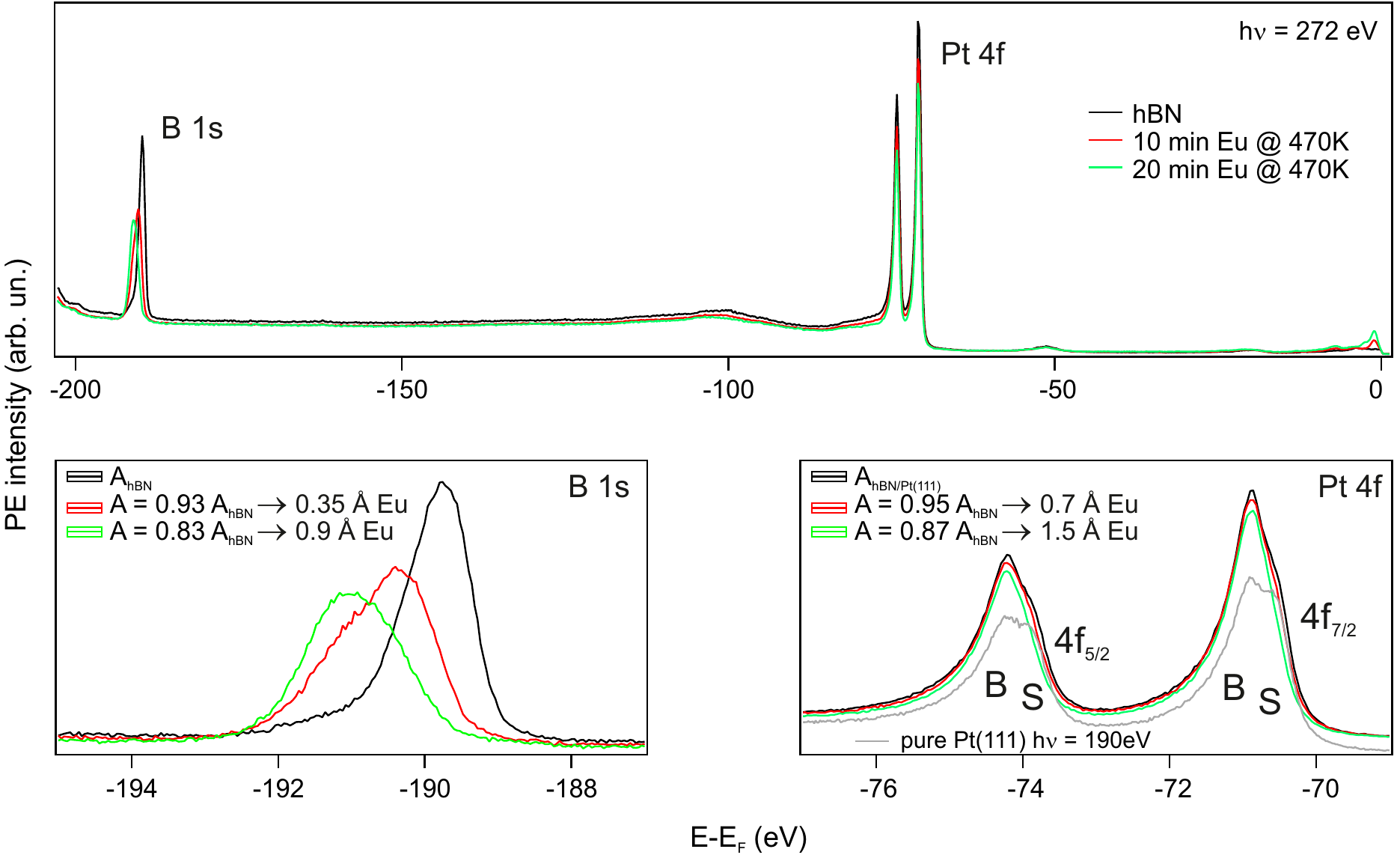}}
\caption{\textbf{XPS analysis for sub-monolayer Eu deposition on hBN/Pt(111) at $T$ = 470K.} XPS survey, B 1s and Pt 4f emissions of 0.7 and 1.5\AA ~Eu coverage on hBN/Pt(111), respectively, taken for h$\nu$ = 272eV. In the case of the Pt 4f also the clean Pt emission is included to better observe the surface and bulk contributions (h$\nu$ = 190eV). After low binding energy normalization, the areas $A$ below the emissions were extracted and from them the overlayer amount was calculated by taking into account the electron mean free path of the electrons from the standard curve~\cite{dench79}. From this analysis we observe that not all Eu is intercalated below the hBN layer but rather stay on top.}
\label{FigS1}
\end{figure}
Fig.~\ref{FigS1} represents X-ray photoemission data from BACH beamline taken at a photon energy $h\nu$ = 272eV. The substrate temperature during Eu deposition for the data shown in Fig.~\ref{FigS1} was $T$ = 470K.
The data set includes measurements of hBN/Pt(111) and two successive Eu intercalations, each for 10 min at a very low rate. The total Eu thickness was calculated from the Pt 4f intensity loss taking into account the mean free path of the electrons. The latter was extracted from the universal curve of the electron mean free path~\cite{dench79} and resulted in Eu thickness of 0.7 and 1.5\AA\, for 10 and 20 min Eu deposition, respectively. In Fig.~\ref{FigS1} one observes the spin orbit split 4f$_{7/2}$ and 4f$_{5/2}$ components of the Pt 4f core level. These are further split in the surface (S) and bulk (B) emissions indicated in the Figure. It is
\begin{figure}[tbph!]
\centerline{\includegraphics[width=0.75\textwidth,angle=0,clip]{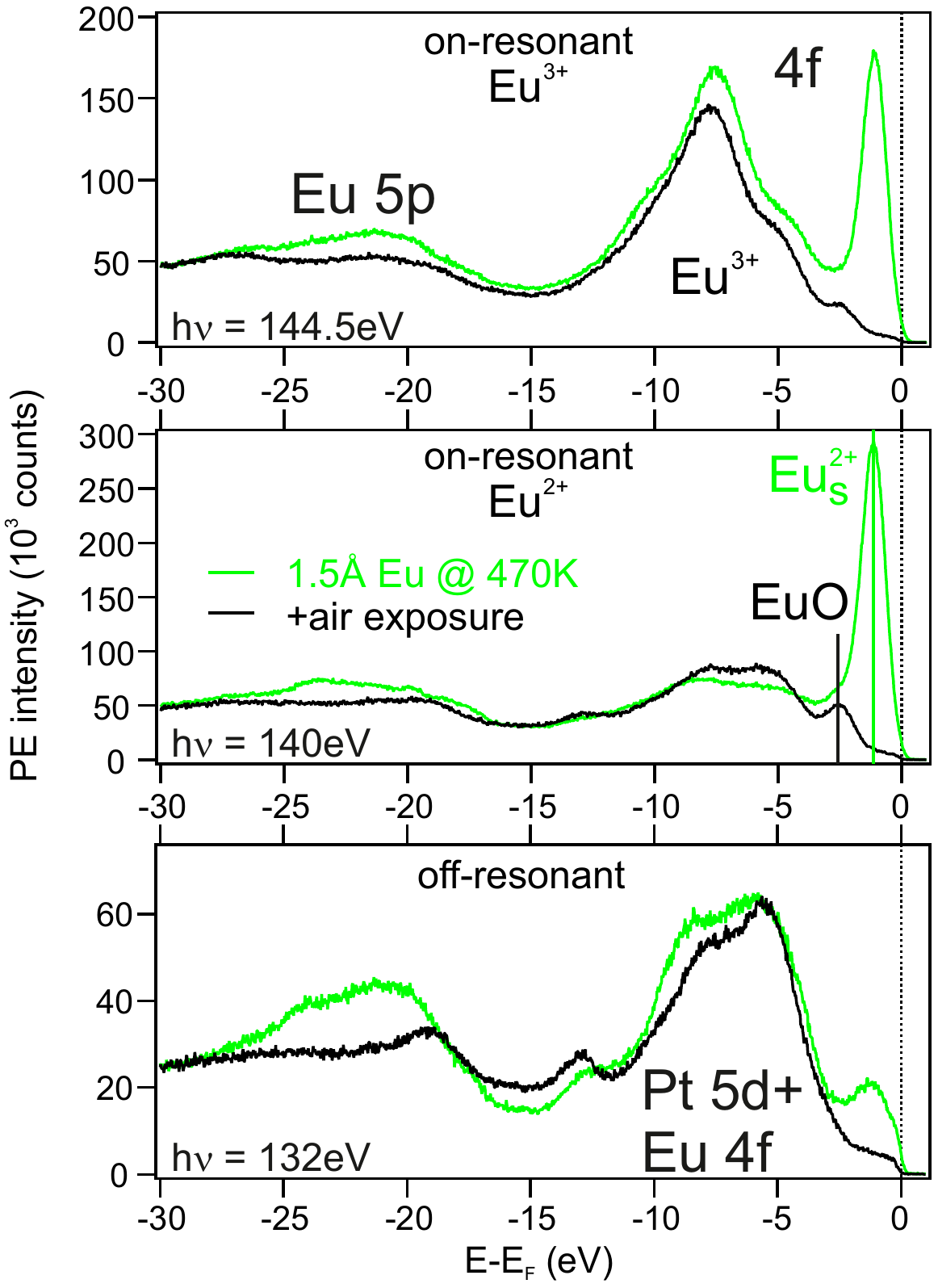}}
\caption{\textbf{Resonant Photoemission at the Eu 4d$\rightarrow$4f absorption edge.} Normal emission photoemission spectra of the valence band region of 1.5\AA~Eu deposited onto a 470K hot hBN/Pt(111) surface for three photon energies corresponding to off, on-resonant Eu$^{2+}$, and on-resonant Eu$^{3+}$ at h$\nu$ = 132, 140, and 144.5eV, respectively. Shown are spectra prior and after exposure of the sample to ambient conditions (5 min, room temperature).}
\label{FigS2}
\end{figure}
interesting to note that the surface emission can be still observed for hBN/Pt(111). This is possible due to the weak interaction of hBN and the Pt, specially at the (111) face that hosts a hBN Moire lattice. After Eu deposition the surface component shrinks much more than the bulk component, revealing that partial Eu intercalation occurred. We again calculate the Eu overlayer thickness but now from the B 1s core level that give us the amount of Eu on top of the hBN layer. The thickness that results from the area diminution is 0.35 and 0.9\AA, respectively, for the two evaporations. This means that during the first deposition 0.35\AA~intercalated and after the second deposition 0.6\AA~out of the 1.5\AA~Eu intercalated. We also observe the doping effect of Eu towards the hBN layer that shifts the B 1s core level to higher binding energies in a similar way as for complete intercalation, see Fig. 6(a) of the main text.

The analysis of the valency has been carried out with resonant photoemission applying photon energies around the 4d$\rightarrow$4f absorption edge. The results can be found in Fig.~\ref{FigS2}. At off-resonant photoemission (h$\nu$ = 132eV) the Eu 4f emission is strongly suppressed, therefore the valence band is dominated by Pt 5d valence band emission. Nevertheless, there is already some Eu 4f emission. The 4f emissions are much better observed for the resonant photon energies h$\nu$ = 140eV and 144.5eV corresponding to Eu$^{2+}$ and Eu$^{3+}$ resonances, respectively~\cite{Schneider1983_RPB28,ColonSantana2012_JPCM,Banik2012_PRB}. The di-valent emission is located at approx. 1.1eV binding energy, the tri-valent multiplet is observed between 4 and 12eV. Due to the unknown cross-section effects around the resonances an exact Eu$^{2+}$/Eu$^{3+}$ ratio cannot be extracted. The tri-valent contribution arises from a part of the intercalated Eu atoms that diffuse further inside the Pt bulk, but can also arise from oxidation of Eu to tri-valent Eu in Eu$_2$O$_3$ of the Eu atoms atop of the hBN layer. The latter oxidation is very difficult to avoid due to the strong reactivity of Eu even under very good ultra-high vacuum conditions~\cite{Schumacher2014_PRB}.
After air exposure, the situation changes drastically. There is still a small di-valent signal, but now at a binding energy of 2.3eV. This Eu$^{2+}$ emission cannot arise from di-valent Eu surrounded by a metallic environment whose binding energy is always lower than 2eV. But an emission at such 2.3eV arises from divalent Eu in EuO at surfaces~\cite{Caspers2011_PRB,ColonSantana2012_JPCM}. This oxide is usually unstable under ambient conditions but seems to be able to form and remain stable under used experimental conditions.
EuO is created either on the surface or at the interface under hBN (and stabilized thanks to the interface). Further investigations would be necessary to distinguish both possibilities.
In any case, we do not observe a possible divalent EuPt$_2$ structure below the hBN after the air exposure process for Eu deposition at 470K substrate temperature.

\begin{figure}[tb!]
\centerline{\includegraphics[width=0.75\textwidth,angle=0,clip]{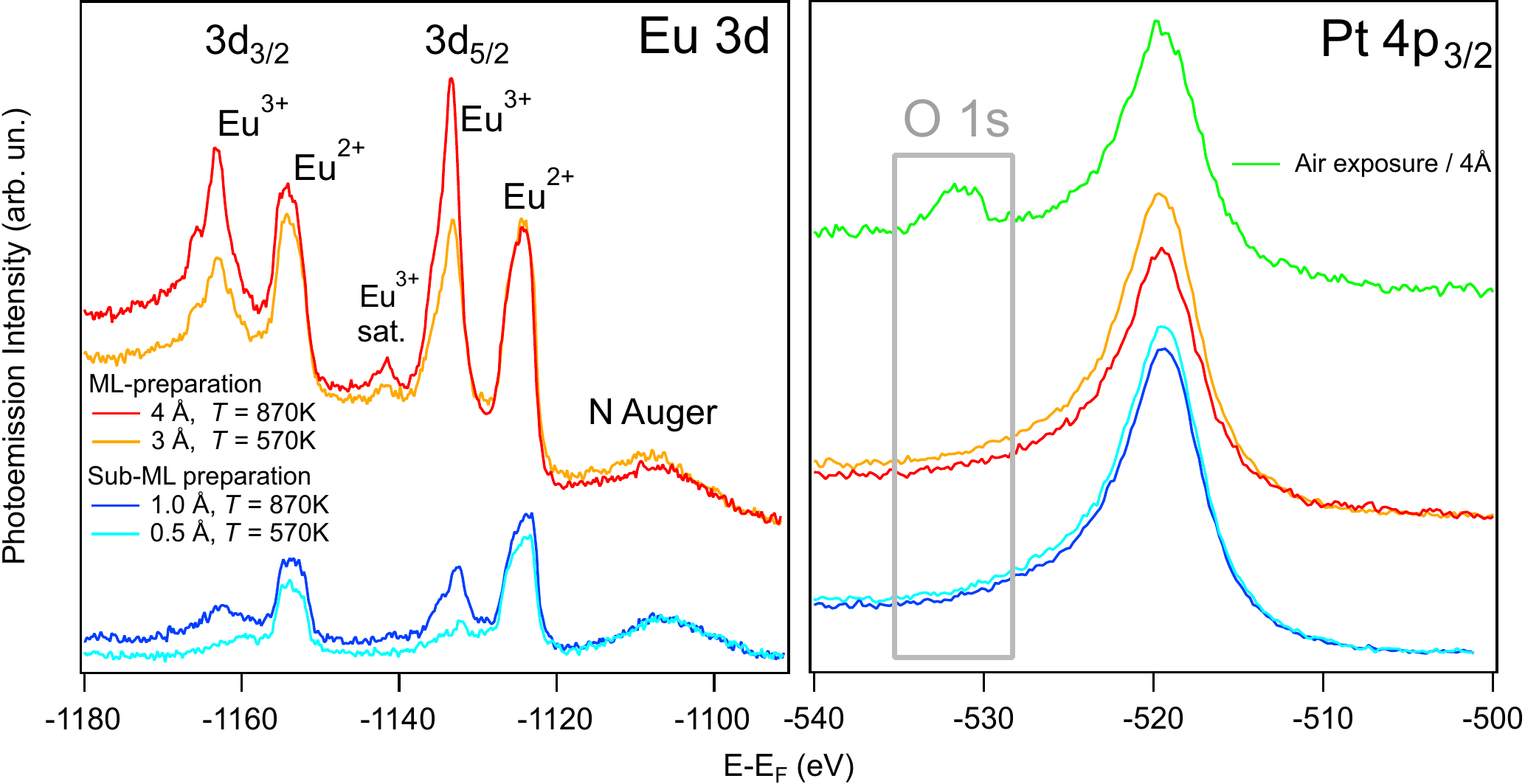}}
\caption{\textbf{Additional X-ray photoemission spectra.} Eu 3d, Pt 4p$_{3/2}$ and O 1s emissions of the XPS series shown in Fig. 3 of the main text.}
\label{additionalXPS_fig3}
\end{figure}

\begin{figure}[tb!]
\centerline{\includegraphics[width=0.75\textwidth,angle=0,clip]{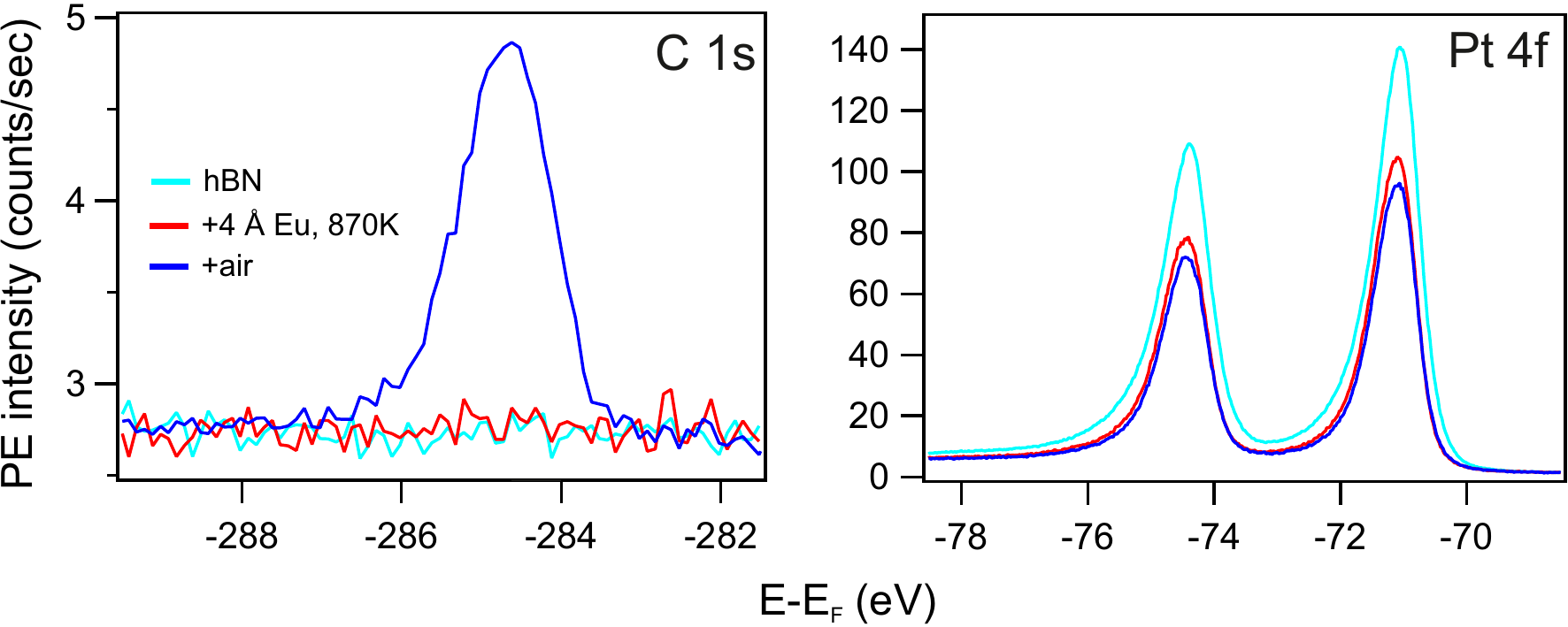}}
\caption{\textbf{Additional X-ray photoemission spectra.} C 1s and Pt 4f emissions of the XPS series shown in Fig. 6(a) of the main text.}
\label{additionalXPS_fig6}
\end{figure}
\section{Additional XPS data for Eu intercalation and air exposure}
\begin{figure}[tb!]
\centerline{\includegraphics[width=0.75\textwidth,angle=0,clip]{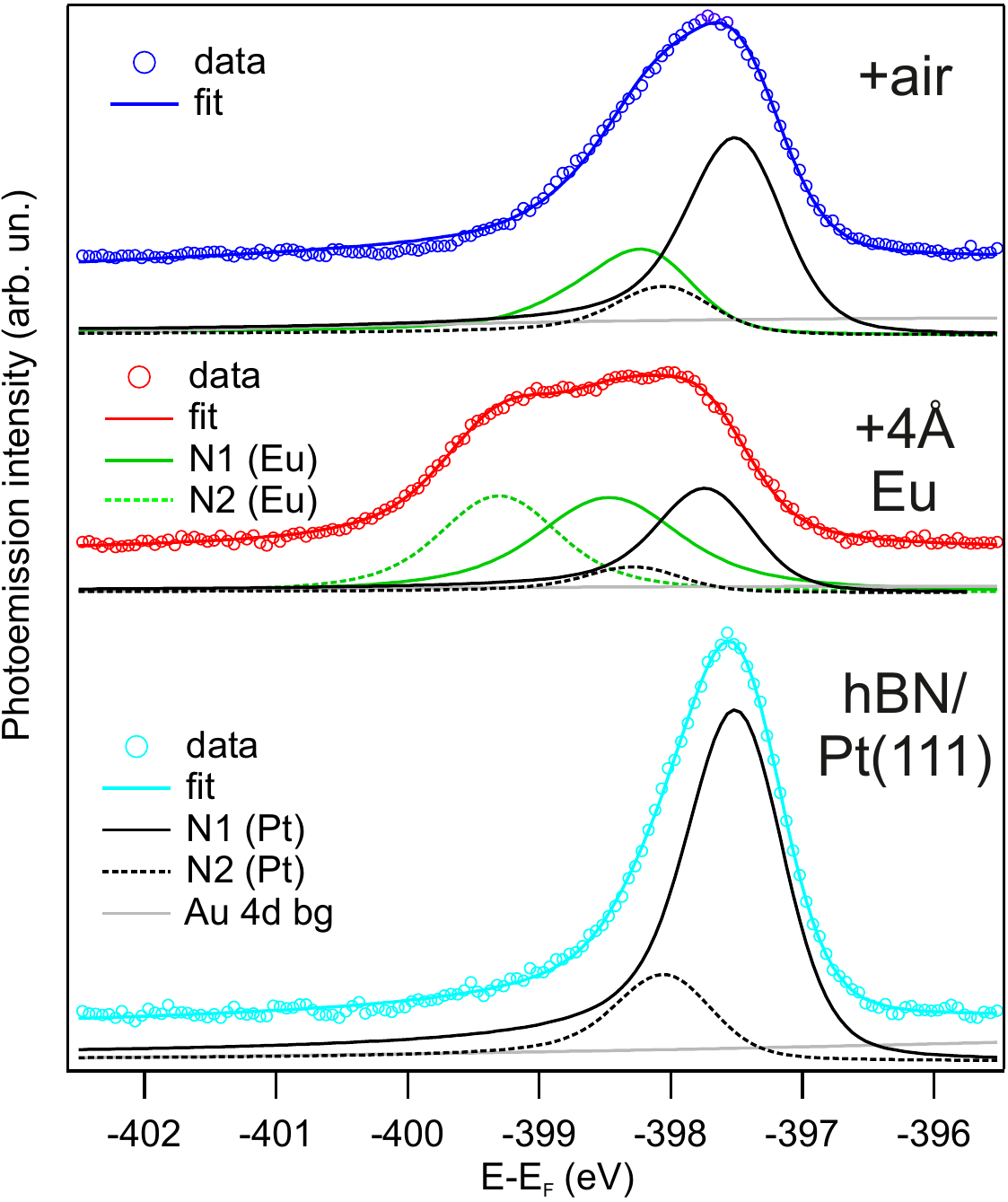}}
\caption{\textbf{Additional X-ray photoemission spectra.} N 1s core level fits for pure hBN/Pt(111), with 4\AA\ Eu intercalated at 870K and after air exposure and annealing. The fits correspond to the data presented in Fig. 6(a) of the main text.}
\label{additionalXPS_fig7}
\end{figure}

{\red{In Figs.~\ref{additionalXPS_fig3}, ~\ref{additionalXPS_fig6}, and ~\ref{additionalXPS_fig7} we present additional XPS spectra for the preparations corresponding to Fig. 3 and 6(a) of the main text, respectively. In Fig.~\ref{additionalXPS_fig3} the Eu 3d spectra of the main text is repeated with its corresponding O 1s emissions. Aim of this figure is to verify that the tri-valent emissions of the Eu 3d core level are not associated to a possible formation of the Europium oxide Eu$_2$O$_3$ since no O 1s emissions are observed prior to air exposure. This is indicated in the second panel of the figure that combines the Pt 3p$_{3/2}$ and O 1s emissions that are close in energy but easily separable. For comparison, we add the spectra of the 4\AA\ Eu spectrum after air exposure to indicate the energy range of O 1s emissions due to the oxide and hydroxide formation that happen at ambient pressure.}}

Furthermore, in Figs.~\ref{additionalXPS_fig6} and ~\ref{additionalXPS_fig7} the C 1s and Pt 4f emissions as well as peak fit results of the N 1s core level spectra of Fig. 6(a) of the main text are shown. Eu deposition and additional air exposure cause that the Pt 4f core level reduces accordingly. After air exposure, an additional C 1s emission is observed that is assigned to adventitious C 1s (saturated hydrocarbons)~\cite{Barr1995_JVacSciTech}. From the intensity loss of the N Auger and Pt 4f emissions and by taking into account the cross sections and surface sensitivity of the XPS spectrum we estimate the hydrocarbon overlayer thickness to roughly 0.5nm after air exposure and annealing. This damping also affects the Eu$^{2+}$ emission that lowered to 1/3 of its original intensity but by taking into account the mentioned damping means that only half of the original Eu$^{2+}$ signal is lost.

{\red{Additionally, the N 1s photoemission was further considered by peak fit procedures and the results are presented in Fig.~\ref{additionalXPS_fig7}. For this purpose, Doniach-\v{S}unjic lineshapes have been employed and a linear background that accounts for the secondary electron emissions from the Au 4d peaks nearby. The leading peak position was observed at 397.2eV binding energy by Preobrajenski et al. \cite{PREOBRAJENSKI2007_PRB}, at 397.3eV by Nagashima et al. \cite{Nagashima1995_PRL}, and we obtain 397.4eV. In a second step we fit the N 1s spectrum after Eu intercalation. We see that in this case we can fit rather well the spectrum with 3 emissions, one corresponding to the doublet found for the clean case, and two additional Doniach-\v{S}unjic lines at binding energy positions 398.5eV and 399.3eV with a rather similar width and intensity. The latter peak positions and splitting resamples the situation for hBN/Rh(111) with positions 397.9eV and 398.6eV \cite{PREOBRAJENSKI2007_PRB} but shifted now by approx. 0.6eV to higher binding energy. The hBN/Rh(111) system - called nanomesh - has a stronger interaction as hBN/Pt(111) leading to a core level shift to higher binding energy and to a much stronger pronunciation of the two peaks that are assigned to emissions from nitrogen atoms with a stronger (high binding energy) and weaker (lower binding energy peak) interaction with the Rh substrate. This situation results from the position of the N atom inside the nanomesh caused by the lattice mismatch between hBN and Rh. In the here considered case of hBN/EuPt$_2$ we observe an even stronger interaction and therefore obtain an even larger binding energy and an even stronger second peak emission. From the XPS fit we also observe that approx. 1/3 of the sample is still Eu free. This means that intercalation, although completed from the point of view that all Eu is below the hBN is not enough to get an alloy everywhere at the interface. After exposure to air, the picture change, Eu gets partially oxidized, partially the di-valent character is preserved. This behaviour results in the disappearance of the very strong interaction peak, a similar less interacting nitrogen peak on Eu, and an increase of the very weak interacting peak that was found initially for hBN/Pt(111). hBN on oxidized europium is very likely to interact weakly in a similar way as hBN on Pt, therefore the latter emission may include such hBN areas on europium oxide.}}

\begin{figure}[btp!]
\centerline{\includegraphics[width=0.5\textwidth,angle=0,clip]{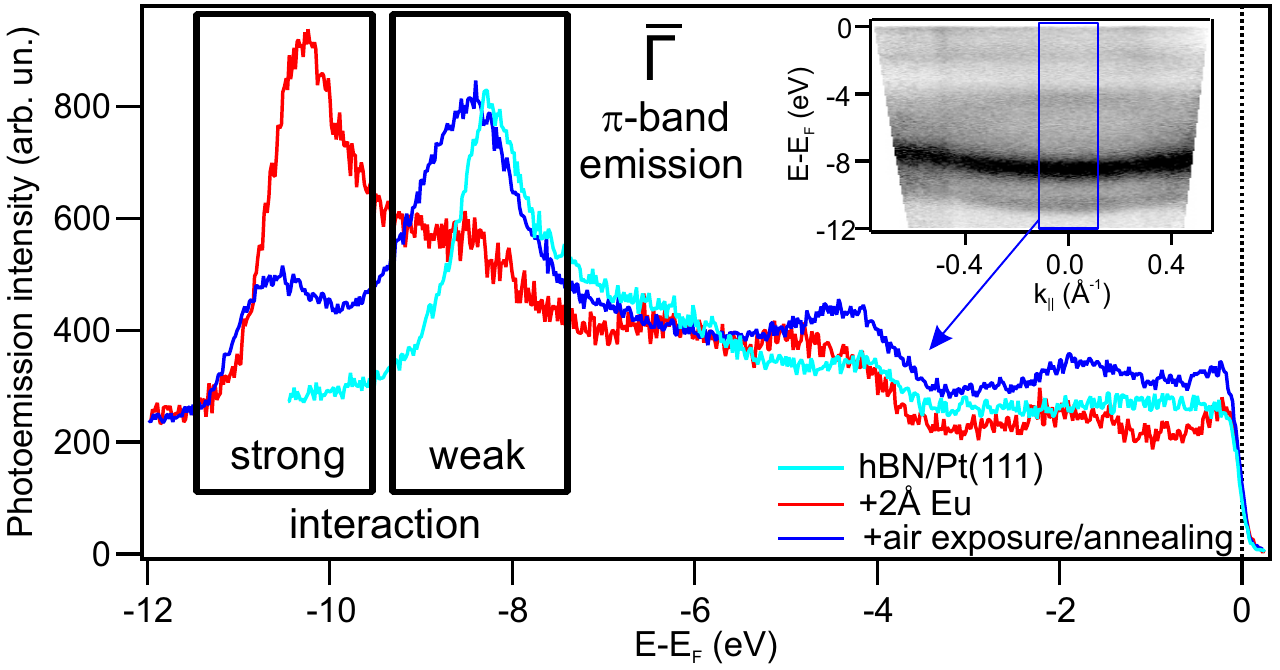}}
\caption{\textbf{Additional valence-band photoemission results.} $\pi$-band emission of hBN at the $\bar\Gamma$-point of the surface Brillouin zone of clean hBN/Pt(111), 4\AA\ Eu intercalated below hBN at 870K, and after air exposure and successive annealing, respectively. Photon energy of the measurement was 40.8eV (HeII emission), measurement temperature 300K. The inset shows the ARPES intensity close to the $\bar\Gamma$-point along $\bar M\bar\Gamma\bar M$ direction after air exposure.}
\label{pi_band_at_Gamma}
\end{figure}

{\red{\section{Additional ARPES data for Eu intercalation and air exposure}
To further observe the influence of the air exposure in ARPES data, in Fig.~\ref{pi_band_at_Gamma} the angle-resolved photoemission spectra at the $\bar\Gamma$-point of the surface Brillouin zone is presented for clean hBN/Pt(111), 4\AA\ Eu intercalated below hBN at 870K, and after air exposure and successive annealing. For better statistics the angular integration range has been set to $\pm$3$^\circ$. Initially, the $\pi$-band for hBN/Pt(111) is found 8.3eV below the Fermi level. Intercalation of 4\AA\ Eu leads to a nearly complete disappearance of the hBN/Pt(111) $\pi$-band and the appearance of a new $\pi$-band at 10.3eV below the Fermi level corresponding to a strongly interacting hBN layer on EuPt$_2$. Air exposure and further annealing to remove the rest-air contamination change again the $\pi$ band structure. One now observes two $\pi$-bands (see inset for dispersion along $\bar M\bar\Gamma\bar M$ direction), the one at higher binding energy has lost intensity and is again attributed to the hBN on the remaining not-oxidized EuPt$_2$ patches while the lower binding energy branch is interpreted as a superposition of bands arising from hBN on the oxidized part and from clean Pt(111) areas, where the Eu intercalation did not take place. The band minimum of the weakly interacting $\pi$-band is now at 8.5eV below the Fermi level. This means that the energy positions of the $\pi$-band on hBN/Pt and on the oxide part are slightly different but too close to be able to be separated in our measurements.}}

\begin{figure}[tbp!]
\centerline{\includegraphics[width=0.5\textwidth,angle=0,clip]{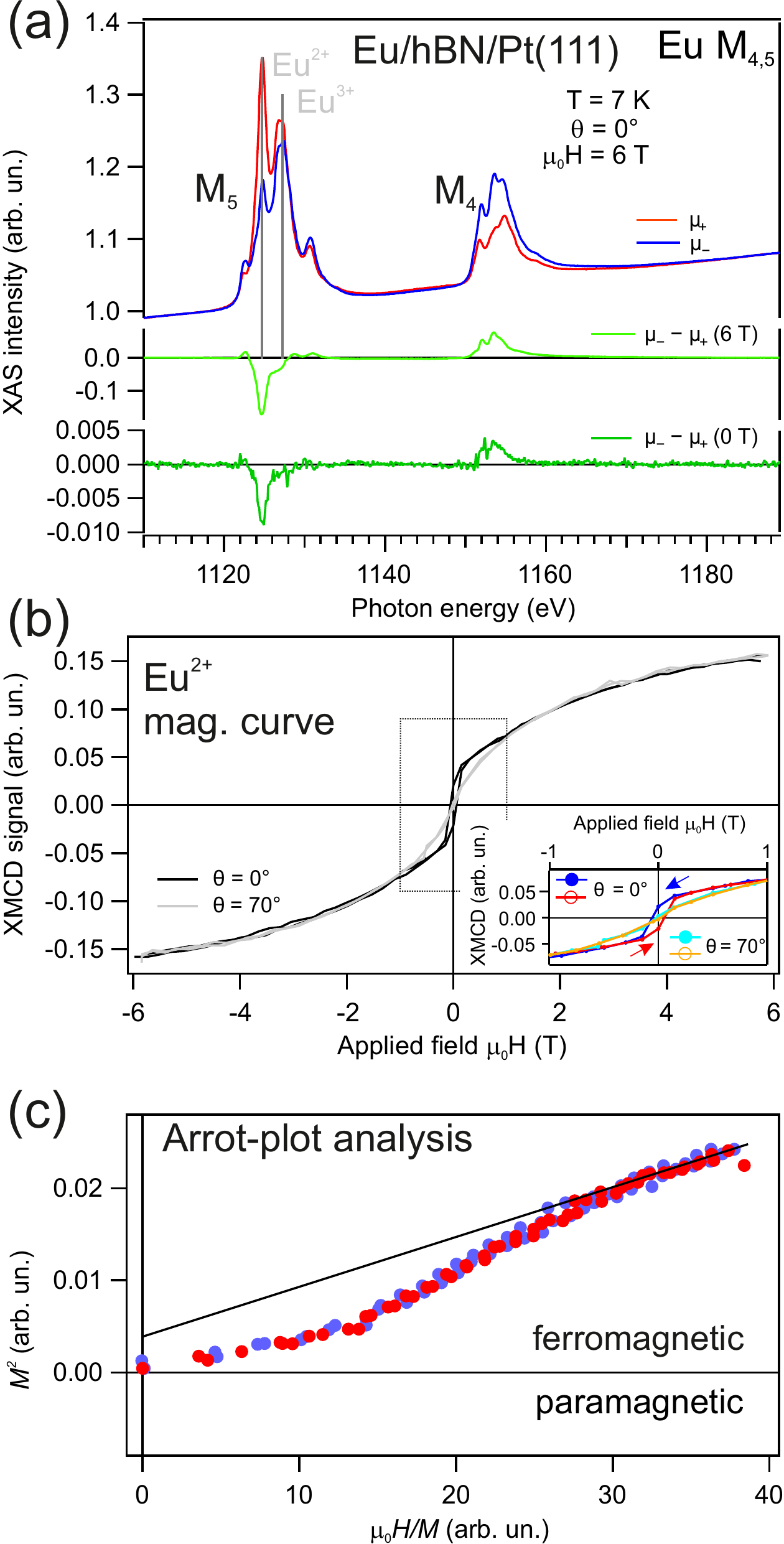}}
\caption{\textbf{Additional XMCD results of Eu.} (a) XAS and XMCD at the Eu M$_{4,5}$ absorption edge in normal incidence geometry. The XMCD difference spectra for 6T and at remanence (applied field 0T) are shown. (b) Eu magnetization curves normal and at $\theta$ = 70$^\circ$ with respect to the field. The inset shows a closeup at small fields to distinguish better the two curves. One clearly observes the more rectangular like shape for the out-of-plane geometry revealing an out-of-plane easy axis. (c) Arrot plot analysis~\cite{Arrott1957_PR} of the normal incidence geometry magnetization curve. The black line presents a linear fit to the high-field (high $\mu_0 H/M$ values).}
\label{mag_loops}
\end{figure}
\section{Additional magnetization measurements, magnetic anisotropy determination}
In Fig.~\ref{mag_loops}(a) we repeat the X-ray absorption spectra of Fig. 5(a) of the main text but add now the XMCD spectrum after removing the applied field keeping the sample in remanence. Although the XMCD signal has dropped considerably (to 4\% of the 6T spectrum), the remanence XMCD spectrum still reveal the same shape as for the applied field. This is proof of the ferromagnetic behaviour of the EuPt$_2$ magnetic layer.

The magnetic anisotropy of the intercalated Eu can be determined from the two different geometries applied. The magnetization curves for the sample perpendicular and at an angle of 70$^\circ$ with respect to the applied field is shown in Fig.~\ref{mag_loops}(b). The close-up at small fields reveal a more rectangular shape of the out-of-plane geometry magnetization loop pointing to this direction as the easy axis of magnetization. The XMCD values close to zero-field has been taken from individual complete XMCD measurements to avoid the typical zero-field artifacts of XMCD measurements.

Fig.~\ref{mag_loops}(c) reveal the Arrot plot analysis~\cite{Arrott1957_PR} of the normal incidence magnetization curve (applied field normal to the sample). This analysis confirms the ferromagnetic state of the EuPt$_2$ surface alloy. In such analysis, the square of the magnetization $M^2$ is plotted against the applied field divided by the magnetization $\mu_0 H/M$. A linear fit of the high field (high $\mu_0 H/M$) values indicates ferromagnetic state if the line hits the ordinate at a positive $M^2$ value, and paramagnetism if the ordinate crossing is negative. For the fit the XMCD values between +5T and +6T and between -5T and -6T have been used.

\pagebreak

%\bibliographystyle{iop_articleTitles}
%\bibliography{../bib}